\documentclass[reqno,12pt]{article}
\usepackage[english]{babel}
\usepackage{cite}
\usepackage{float}
\usepackage{ae} 
\usepackage[T1]{fontenc}
\usepackage{amsmath}
\numberwithin{equation}{section}

\usepackage{amssymb}
\usepackage{mathrsfs}
\usepackage{graphicx}
\usepackage{mdframed}
 \usepackage{mathrsfs}
\usepackage{amsfonts,mathtools}
\usepackage{color}
\definecolor{darkblue}{cmyk}{0.9,0.9,0,0}
\usepackage[colorlinks=true,linkcolor=darkblue,citecolor=darkblue,urlcolor=darkblue]{hyperref}
\usepackage{epsfig}

\usepackage{graphicx}
\usepackage{tikz}

\usetikzlibrary{fadings}
\usetikzlibrary{patterns}
\usetikzlibrary{shadows.blur}
\usetikzlibrary{shapes}

\tikzset{
	pattern size/.store in=\mcSize, 
	pattern size = 5pt,
	pattern thickness/.store in=\mcThickness, 
	pattern thickness = 0.3pt,
	pattern radius/.store in=\mcRadius, 
	pattern radius = 1pt}
\makeatletter
\pgfutil@ifundefined{pgf@pattern@name@_tfrboppns}{
	\makeatletter
	\pgfdeclarepatternformonly[\mcRadius,\mcThickness,\mcSize]{_tfrboppns}
	{\pgfpoint{-0.5*\mcSize}{-0.5*\mcSize}}
	{\pgfpoint{0.5*\mcSize}{0.5*\mcSize}}
	{\pgfpoint{\mcSize}{\mcSize}}
	{
		\pgfsetcolor{\tikz@pattern@color}
		\pgfsetlinewidth{\mcThickness}
		\pgfpathcircle\pgfpointorigin{\mcRadius}
		\pgfusepath{stroke}
}}
\makeatother

\tikzset{
pattern size/.store in=\mcSize, 
pattern size = 5pt,
pattern thickness/.store in=\mcThickness, 
pattern thickness = 0.3pt,
pattern radius/.store in=\mcRadius, 
pattern radius = 1pt}
\makeatletter
\pgfutil@ifundefined{pgf@pattern@_7q7remqtl}{
\pgfdeclarepatternformonly[\mcThickness,\mcSize]{_7q7remqtl}
{\pgfpointorigin}
{\pgfpoint{\mcSize+\mcThickness}{\mcSize+\mcThickness}}
{\pgfpoint{\mcSize}{\mcSize}}{
\pgfsetcolor{\tikz@pattern@color}
\pgfsetlinewidth{\mcThickness}
\pgfpathmoveto{\pgfpointorigin}
\pgfpathlineto{\pgfpoint{0pt}{0.5*\mcSize}}
\pgfpathlineto{\pgfpoint{\mcSize}{0.5*\mcSize}}
\pgfpathmoveto{\pgfpoint{0.5*\mcSize}{0.5*\mcSize}}
\pgfpathlineto{\pgfpoint{0.5*\mcSize}{\mcSize}}
\pgfpathmoveto{\pgfpoint{0pt}{\mcSize}}
\pgfpathlineto{\pgfpoint{\mcSize}{\mcSize}}
\pgfusepath{stroke}}}
\makeatother

\tikzset{
	pattern size/.store in=\mcSize, 
	pattern size = 5pt,
	pattern thickness/.store in=\mcThickness, 
	pattern thickness = 0.3pt,
	pattern radius/.store in=\mcRadius, 
	pattern radius = 1pt}
\makeatletter
\pgfutil@ifundefined{pgf@pattern@name@_eine17d0l}{
	\makeatletter
	\pgfdeclarepatternformonly[\mcRadius,\mcThickness,\mcSize]{_eine17d0l}
	{\pgfpoint{-0.5*\mcSize}{-0.5*\mcSize}}
	{\pgfpoint{0.5*\mcSize}{0.5*\mcSize}}
	{\pgfpoint{\mcSize}{\mcSize}}
	{
		\pgfsetcolor{\tikz@pattern@color}
		\pgfsetlinewidth{\mcThickness}
		\pgfpathcircle\pgfpointorigin{\mcRadius}
		\pgfusepath{stroke}
}}
\makeatother

\tikzset{
pattern size/.store in=\mcSize, 
pattern size = 5pt,
pattern thickness/.store in=\mcThickness, 
pattern thickness = 0.3pt,
pattern radius/.store in=\mcRadius, 
pattern radius = 1pt}
\makeatletter
\pgfutil@ifundefined{pgf@pattern@_yyxs2f8x9}{
\pgfdeclarepatternformonly[\mcThickness,\mcSize]{_yyxs2f8x9}
{\pgfpointorigin}
{\pgfpoint{\mcSize+\mcThickness}{\mcSize+\mcThickness}}
{\pgfpoint{\mcSize}{\mcSize}}{
\pgfsetcolor{\tikz@pattern@color}
\pgfsetlinewidth{\mcThickness}
\pgfpathmoveto{\pgfpointorigin}
\pgfpathlineto{\pgfpoint{0pt}{0.5*\mcSize}}
\pgfpathlineto{\pgfpoint{\mcSize}{0.5*\mcSize}}
\pgfpathmoveto{\pgfpoint{0.5*\mcSize}{0.5*\mcSize}}
\pgfpathlineto{\pgfpoint{0.5*\mcSize}{\mcSize}}
\pgfpathmoveto{\pgfpoint{0pt}{\mcSize}}
\pgfpathlineto{\pgfpoint{\mcSize}{\mcSize}}
\pgfusepath{stroke}}}
\makeatother
 
\tikzset{
pattern size/.store in=\mcSize, 
pattern size = 5pt,
pattern thickness/.store in=\mcThickness, 
pattern thickness = 0.3pt,
pattern radius/.store in=\mcRadius, 
pattern radius = 1pt}
\makeatletter
\pgfutil@ifundefined{pgf@pattern@_bixux0kyf}{
\pgfdeclarepatternformonly[\mcThickness,\mcSize]{_bixux0kyf}
{\pgfpointorigin}
{\pgfpoint{\mcSize+\mcThickness}{\mcSize+\mcThickness}}
{\pgfpoint{\mcSize}{\mcSize}}{
\pgfsetcolor{\tikz@pattern@color}
\pgfsetlinewidth{\mcThickness}
\pgfpathmoveto{\pgfpointorigin}
\pgfpathlineto{\pgfpoint{0pt}{0.5*\mcSize}}
\pgfpathlineto{\pgfpoint{\mcSize}{0.5*\mcSize}}
\pgfpathmoveto{\pgfpoint{0.5*\mcSize}{0.5*\mcSize}}
\pgfpathlineto{\pgfpoint{0.5*\mcSize}{\mcSize}}
\pgfpathmoveto{\pgfpoint{0pt}{\mcSize}}
\pgfpathlineto{\pgfpoint{\mcSize}{\mcSize}}
\pgfusepath{stroke}}}
\makeatother

\tikzset{
	pattern size/.store in=\mcSize, 
	pattern size = 5pt,
	pattern thickness/.store in=\mcThickness, 
	pattern thickness = 0.3pt,
	pattern radius/.store in=\mcRadius, 
	pattern radius = 1pt}
\makeatletter
\pgfutil@ifundefined{pgf@pattern@_1uqk3jyyv}{
	\pgfdeclarepatternformonly[\mcThickness,\mcSize]{_1uqk3jyyv}
	{\pgfpointorigin}
	{\pgfpoint{\mcSize+\mcThickness}{\mcSize+\mcThickness}}
	{\pgfpoint{\mcSize}{\mcSize}}{
		\pgfsetcolor{\tikz@pattern@color}
		\pgfsetlinewidth{\mcThickness}
		\pgfpathmoveto{\pgfpointorigin}
		\pgfpathlineto{\pgfpoint{0pt}{0.5*\mcSize}}
		\pgfpathlineto{\pgfpoint{\mcSize}{0.5*\mcSize}}
		\pgfpathmoveto{\pgfpoint{0.5*\mcSize}{0.5*\mcSize}}
		\pgfpathlineto{\pgfpoint{0.5*\mcSize}{\mcSize}}
		\pgfpathmoveto{\pgfpoint{0pt}{\mcSize}}
		\pgfpathlineto{\pgfpoint{\mcSize}{\mcSize}}
		\pgfusepath{stroke}}}
\makeatother
\tikzset{every picture/.style={line width=0.75pt}} 

\tikzset{
pattern size/.store in=\mcSize, 
pattern size = 5pt,
pattern thickness/.store in=\mcThickness, 
pattern thickness = 0.3pt,
pattern radius/.store in=\mcRadius, 
pattern radius = 1pt}
\makeatletter
\pgfutil@ifundefined{pgf@pattern@_tkc5crui7}{
\pgfdeclarepatternformonly[\mcThickness,\mcSize]{_tkc5crui7}
{\pgfpointorigin}
{\pgfpoint{\mcSize+\mcThickness}{\mcSize+\mcThickness}}
{\pgfpoint{\mcSize}{\mcSize}}{
\pgfsetcolor{\tikz@pattern@color}
\pgfsetlinewidth{\mcThickness}
\pgfpathmoveto{\pgfpointorigin}
\pgfpathlineto{\pgfpoint{0pt}{0.5*\mcSize}}
\pgfpathlineto{\pgfpoint{\mcSize}{0.5*\mcSize}}
\pgfpathmoveto{\pgfpoint{0.5*\mcSize}{0.5*\mcSize}}
\pgfpathlineto{\pgfpoint{0.5*\mcSize}{\mcSize}}
\pgfpathmoveto{\pgfpoint{0pt}{\mcSize}}
\pgfpathlineto{\pgfpoint{\mcSize}{\mcSize}}
\pgfusepath{stroke}}}
\makeatother

\tikzset{
pattern size/.store in=\mcSize, 
pattern size = 5pt,
pattern thickness/.store in=\mcThickness, 
pattern thickness = 0.3pt,
pattern radius/.store in=\mcRadius, 
pattern radius = 1pt}
\makeatletter
\pgfutil@ifundefined{pgf@pattern@name@_7iz9i42p8}{
\makeatletter
\pgfdeclarepatternformonly[\mcRadius,\mcThickness,\mcSize]{_7iz9i42p8}
{\pgfpoint{-0.5*\mcSize}{-0.5*\mcSize}}
{\pgfpoint{0.5*\mcSize}{0.5*\mcSize}}
{\pgfpoint{\mcSize}{\mcSize}}
{
\pgfsetcolor{\tikz@pattern@color}
\pgfsetlinewidth{\mcThickness}
\pgfpathcircle\pgfpointorigin{\mcRadius}
\pgfusepath{stroke}
}}
\makeatother

\tikzset{
pattern size/.store in=\mcSize, 
pattern size = 5pt,
pattern thickness/.store in=\mcThickness, 
pattern thickness = 0.3pt,
pattern radius/.store in=\mcRadius, 
pattern radius = 1pt}
\makeatletter
\pgfutil@ifundefined{pgf@pattern@_cdw6nuggu}{
\pgfdeclarepatternformonly[\mcThickness,\mcSize]{_cdw6nuggu}
{\pgfpointorigin}
{\pgfpoint{\mcSize+\mcThickness}{\mcSize+\mcThickness}}
{\pgfpoint{\mcSize}{\mcSize}}{
\pgfsetcolor{\tikz@pattern@color}
\pgfsetlinewidth{\mcThickness}
\pgfpathmoveto{\pgfpointorigin}
\pgfpathlineto{\pgfpoint{0pt}{0.5*\mcSize}}
\pgfpathlineto{\pgfpoint{\mcSize}{0.5*\mcSize}}
\pgfpathmoveto{\pgfpoint{0.5*\mcSize}{0.5*\mcSize}}
\pgfpathlineto{\pgfpoint{0.5*\mcSize}{\mcSize}}
\pgfpathmoveto{\pgfpoint{0pt}{\mcSize}}
\pgfpathlineto{\pgfpoint{\mcSize}{\mcSize}}
\pgfusepath{stroke}}}
\makeatother

\tikzset{
pattern size/.store in=\mcSize, 
pattern size = 5pt,
pattern thickness/.store in=\mcThickness, 
pattern thickness = 0.3pt,
pattern radius/.store in=\mcRadius, 
pattern radius = 1pt}
\makeatletter
\pgfutil@ifundefined{pgf@pattern@name@_8iavq2zkj}{
\makeatletter
\pgfdeclarepatternformonly[\mcRadius,\mcThickness,\mcSize]{_8iavq2zkj}
{\pgfpoint{-0.5*\mcSize}{-0.5*\mcSize}}
{\pgfpoint{0.5*\mcSize}{0.5*\mcSize}}
{\pgfpoint{\mcSize}{\mcSize}}
{
\pgfsetcolor{\tikz@pattern@color}
\pgfsetlinewidth{\mcThickness}
\pgfpathcircle\pgfpointorigin{\mcRadius}
\pgfusepath{stroke}
}}
\makeatother

\tikzset{
pattern size/.store in=\mcSize, 
pattern size = 5pt,
pattern thickness/.store in=\mcThickness, 
pattern thickness = 0.3pt,
pattern radius/.store in=\mcRadius, 
pattern radius = 1pt}
\makeatletter
\pgfutil@ifundefined{pgf@pattern@name@_hzkc6ogze}{
\makeatletter
\pgfdeclarepatternformonly[\mcRadius,\mcThickness,\mcSize]{_hzkc6ogze}
{\pgfpoint{-0.5*\mcSize}{-0.5*\mcSize}}
{\pgfpoint{0.5*\mcSize}{0.5*\mcSize}}
{\pgfpoint{\mcSize}{\mcSize}}
{
\pgfsetcolor{\tikz@pattern@color}
\pgfsetlinewidth{\mcThickness}
\pgfpathcircle\pgfpointorigin{\mcRadius}
\pgfusepath{stroke}
}}
\makeatother

\tikzset{
pattern size/.store in=\mcSize, 
pattern size = 5pt,
pattern thickness/.store in=\mcThickness, 
pattern thickness = 0.3pt,
pattern radius/.store in=\mcRadius, 
pattern radius = 1pt}
\makeatletter
\pgfutil@ifundefined{pgf@pattern@_6zrpb3n5c}{
\pgfdeclarepatternformonly[\mcThickness,\mcSize]{_6zrpb3n5c}
{\pgfpointorigin}
{\pgfpoint{\mcSize+\mcThickness}{\mcSize+\mcThickness}}
{\pgfpoint{\mcSize}{\mcSize}}{
\pgfsetcolor{\tikz@pattern@color}
\pgfsetlinewidth{\mcThickness}
\pgfpathmoveto{\pgfpointorigin}
\pgfpathlineto{\pgfpoint{0pt}{0.5*\mcSize}}
\pgfpathlineto{\pgfpoint{\mcSize}{0.5*\mcSize}}
\pgfpathmoveto{\pgfpoint{0.5*\mcSize}{0.5*\mcSize}}
\pgfpathlineto{\pgfpoint{0.5*\mcSize}{\mcSize}}
\pgfpathmoveto{\pgfpoint{0pt}{\mcSize}}
\pgfpathlineto{\pgfpoint{\mcSize}{\mcSize}}
\pgfusepath{stroke}}}
\makeatother
 
\tikzset{
pattern size/.store in=\mcSize, 
pattern size = 5pt,
pattern thickness/.store in=\mcThickness, 
pattern thickness = 0.3pt,
pattern radius/.store in=\mcRadius, 
pattern radius = 1pt}
\makeatletter
\pgfutil@ifundefined{pgf@pattern@name@_6ppuhd14k}{
\makeatletter
\pgfdeclarepatternformonly[\mcRadius,\mcThickness,\mcSize]{_6ppuhd14k}
{\pgfpoint{-0.5*\mcSize}{-0.5*\mcSize}}
{\pgfpoint{0.5*\mcSize}{0.5*\mcSize}}
{\pgfpoint{\mcSize}{\mcSize}}
{
\pgfsetcolor{\tikz@pattern@color}
\pgfsetlinewidth{\mcThickness}
\pgfpathcircle\pgfpointorigin{\mcRadius}
\pgfusepath{stroke}
}}
\makeatother

\tikzset{
pattern size/.store in=\mcSize, 
pattern size = 5pt,
pattern thickness/.store in=\mcThickness, 
pattern thickness = 0.3pt,
pattern radius/.store in=\mcRadius, 
pattern radius = 1pt}
\makeatletter
\pgfutil@ifundefined{pgf@pattern@_2sbugfwpz}{
\pgfdeclarepatternformonly[\mcThickness,\mcSize]{_2sbugfwpz}
{\pgfpointorigin}
{\pgfpoint{\mcSize+\mcThickness}{\mcSize+\mcThickness}}
{\pgfpoint{\mcSize}{\mcSize}}{
\pgfsetcolor{\tikz@pattern@color}
\pgfsetlinewidth{\mcThickness}
\pgfpathmoveto{\pgfpointorigin}
\pgfpathlineto{\pgfpoint{0pt}{0.5*\mcSize}}
\pgfpathlineto{\pgfpoint{\mcSize}{0.5*\mcSize}}
\pgfpathmoveto{\pgfpoint{0.5*\mcSize}{0.5*\mcSize}}
\pgfpathlineto{\pgfpoint{0.5*\mcSize}{\mcSize}}
\pgfpathmoveto{\pgfpoint{0pt}{\mcSize}}
\pgfpathlineto{\pgfpoint{\mcSize}{\mcSize}}
\pgfusepath{stroke}}}
\makeatother

\tikzset{
pattern size/.store in=\mcSize, 
pattern size = 5pt,
pattern thickness/.store in=\mcThickness, 
pattern thickness = 0.3pt,
pattern radius/.store in=\mcRadius, 
pattern radius = 1pt}
\makeatletter
\pgfutil@ifundefined{pgf@pattern@_icy489xm3}{
\pgfdeclarepatternformonly[\mcThickness,\mcSize]{_icy489xm3}
{\pgfpointorigin}
{\pgfpoint{\mcSize+\mcThickness}{\mcSize+\mcThickness}}
{\pgfpoint{\mcSize}{\mcSize}}{
\pgfsetcolor{\tikz@pattern@color}
\pgfsetlinewidth{\mcThickness}
\pgfpathmoveto{\pgfpointorigin}
\pgfpathlineto{\pgfpoint{0pt}{0.5*\mcSize}}
\pgfpathlineto{\pgfpoint{\mcSize}{0.5*\mcSize}}
\pgfpathmoveto{\pgfpoint{0.5*\mcSize}{0.5*\mcSize}}
\pgfpathlineto{\pgfpoint{0.5*\mcSize}{\mcSize}}
\pgfpathmoveto{\pgfpoint{0pt}{\mcSize}}
\pgfpathlineto{\pgfpoint{\mcSize}{\mcSize}}
\pgfusepath{stroke}}}
\makeatother

\tikzset{
pattern size/.store in=\mcSize, 
pattern size = 5pt,
pattern thickness/.store in=\mcThickness, 
pattern thickness = 0.3pt,
pattern radius/.store in=\mcRadius, 
pattern radius = 1pt}
\makeatletter
\pgfutil@ifundefined{pgf@pattern@name@_ar6uez0vk}{
\makeatletter
\pgfdeclarepatternformonly[\mcRadius,\mcThickness,\mcSize]{_ar6uez0vk}
{\pgfpoint{-0.5*\mcSize}{-0.5*\mcSize}}
{\pgfpoint{0.5*\mcSize}{0.5*\mcSize}}
{\pgfpoint{\mcSize}{\mcSize}}
{
\pgfsetcolor{\tikz@pattern@color}
\pgfsetlinewidth{\mcThickness}
\pgfpathcircle\pgfpointorigin{\mcRadius}
\pgfusepath{stroke}
}}
\makeatother

\tikzset{
pattern size/.store in=\mcSize, 
pattern size = 5pt,
pattern thickness/.store in=\mcThickness, 
pattern thickness = 0.3pt,
pattern radius/.store in=\mcRadius, 
pattern radius = 1pt}
\makeatletter
\pgfutil@ifundefined{pgf@pattern@name@_bn8zr1qen}{
\makeatletter
\pgfdeclarepatternformonly[\mcRadius,\mcThickness,\mcSize]{_bn8zr1qen}
{\pgfpoint{-0.5*\mcSize}{-0.5*\mcSize}}
{\pgfpoint{0.5*\mcSize}{0.5*\mcSize}}
{\pgfpoint{\mcSize}{\mcSize}}
{
\pgfsetcolor{\tikz@pattern@color}
\pgfsetlinewidth{\mcThickness}
\pgfpathcircle\pgfpointorigin{\mcRadius}
\pgfusepath{stroke}
}}
\makeatother

\tikzset{
pattern size/.store in=\mcSize, 
pattern size = 5pt,
pattern thickness/.store in=\mcThickness, 
pattern thickness = 0.3pt,
pattern radius/.store in=\mcRadius, 
pattern radius = 1pt}
\makeatletter
\pgfutil@ifundefined{pgf@pattern@_3une2mc6d}{
\pgfdeclarepatternformonly[\mcThickness,\mcSize]{_3une2mc6d}
{\pgfpointorigin}
{\pgfpoint{\mcSize+\mcThickness}{\mcSize+\mcThickness}}
{\pgfpoint{\mcSize}{\mcSize}}{
\pgfsetcolor{\tikz@pattern@color}
\pgfsetlinewidth{\mcThickness}
\pgfpathmoveto{\pgfpointorigin}
\pgfpathlineto{\pgfpoint{0pt}{0.5*\mcSize}}
\pgfpathlineto{\pgfpoint{\mcSize}{0.5*\mcSize}}
\pgfpathmoveto{\pgfpoint{0.5*\mcSize}{0.5*\mcSize}}
\pgfpathlineto{\pgfpoint{0.5*\mcSize}{\mcSize}}
\pgfpathmoveto{\pgfpoint{0pt}{\mcSize}}
\pgfpathlineto{\pgfpoint{\mcSize}{\mcSize}}
\pgfusepath{stroke}}}
\makeatother

\newcommand{\beq}{\begin{equation}}
\newcommand{\eeq}{\end{equation}}
\newcommand\beqa{\begin{eqnarray}}
\newcommand\eeqa{\end{eqnarray}}
\newcommand\bea{\begin{array}}
\newcommand\eea{\end{array}}

\usepackage{varioref}
\usepackage{makeidx}
\makeindex

\begin{document}

\thispagestyle{empty}

\renewcommand{\thefootnote}{\fnsymbol{footnote}}
\setcounter{page}{1}
\setcounter{footnote}{0}
\setcounter{figure}{0}

\vspace{-0.4in}

\begin{center}
$$~$$
{\Large
Flat limit of AdS/CFT from AdS geodesics: scattering amplitudes and antipodal matching of Li\'enard-Wiechert fields \\
\par}
\vspace{1.0cm}

\textrm{Sarthak Duary$^\text{\tiny 1}$\footnote{\tt sarthakduary@tsinghua.edu.cn} and Shivam Upadhyay$^\text{\tiny 2}$\footnote{\tt shivam.dep@gmail.com}}

 \vspace{1.2cm}
\footnotesize{
{\it $^\text{\tiny 1}$Yau Mathematical Sciences Center (YMSC), Tsinghua University, Beijing 100084, China}\\
{\it $^\text{\tiny 2}$Chennai Mathematical Institute, Siruseri, SIPCOT, Chennai 603103}\\

\vspace{4mm}
}
\end{center}

\vspace{2mm}
\begin{abstract}
We revisit the flat limit of AdS/CFT from the point of view of geodesics in AdS. We show that the flat space scattering amplitudes can be constructed from operator insertions where the geodesics of the particles corresponding to the operators hit the conformal boundary of AdS. Further, we compute the Li\'enard-Wiechert solutions in AdS by boosting a static charge using AdS isometries and show that the solutions are antipodally matched between two regions, separated by a global time difference of $\Delta\tau=\pi$. Going to the boundary of AdS along null geodesics, in the flat limit, this antipodal matching leads to the flat space antipodal matching near spatial infinity.
\end{abstract}

\newpage

\addtocontents{toc}{\protect\setcounter{tocdepth}{2}}

\setcounter{page}{1}
\renewcommand{\thefootnote}{\arabic{footnote}}
\setcounter{footnote}{0}

{
\tableofcontents
}


\section{Introduction}







Despite significant progress in understanding quantum gravity in asymptotically Anti-de Sitter (AdS) spacetimes, we still have a limited grasp of how to describe gravity holographically in asymptotically flat space. If we consider an observer living in
AdS, they would perceive the space as flat if their observations are limited to scales below the AdS length scale. 
The key takeaway is that flat space is a part of AdS space, meaning that the physics of flat space is inherently encoded in AdS. Since physics in AdS is dual to conformal field theories (CFTs), it is reasonable to conclude that these CFTs must encapsulate the physics of flat space in an additional dimension. The challenge, then, is to determine how to extract the flat space physics encoded in these CFTs. The question is: given that flat space physics is represented in the CFT, how can we explicitly derive flat space observables, like the $\mathcal{S}$-matrix, from the boundary theory? Understanding flat space holography entails a method for computing the flat space $\mathcal{S}$-matrix within a lower-dimensional framework. 
In the flat limit, the boundary correlation functions of certain operators should transform into $\mathcal{S}$-matrix elements. This concept traces back to the early development of the AdS/CFT correspondence \cite{Polchinski:1999ry, Susskind:1998vk, Giddings:1999qu, Giddings:1999jq, Gary:2009mi}. Subsequent works have proposed several potential approaches \cite{Okuda:2010ym, Fitzpatrick:2011jn, Raju:2012zr, Maldacena:2015iua, Paulos:2016fap, Dubovsky:2017cnj}. More recently, the flat limit of AdS/CFT has received renewed attention \cite{Hijano:2019qmi, Hijano:2020szl, Li:2021snj, Duary:2023gqg, Komatsu:2020sag, vanRees:2022zmr, vanRees:2023fcf, Duary:2022pyv, Duary:2022afn, Banerjee:2022oll, Li:2023azu, Li:2022tby, Duary:2024fii}. 

We will briefly outline the construction of the $\mathcal{S}$-matrix using techniques from AdS/CFT. The primary objective of flat space scattering theory is to compute the $\mathcal{S}$-matrix, which involves determining the overlap between two different scattering states of the full Hamiltonian.\footnote{In interacting theories, full Hamiltonian states are often complex, so approximations are used for scattering states. One such approximation is the asymptotic decoupling assumption, which assumes the Hamiltonian at null infinity (the boundary of flat space) behaves as if it were free.} The goal is to construct Fock space scattering states, approximated as free particles at asymptotic infinity, using AdS/CFT techniques. This involves bulk operator reconstruction, where we express a local bulk operator deep within AdS in terms of boundary operators on the CFT. In a flat space scattering region, we apply the asymptotic decoupling approximation, simplifying the field to behave like a free field. This allows us to decompose the field into plane waves, extract creation and annihilation operators via Fourier transform, and establish the mapping between flat space operators and CFT boundary operators. For a detailed computation of the mapping between creation and annihilation operators in flat space and CFT operators, we direct the reader to \cite{Hijano:2019qmi, Hijano:2020szl, Li:2021snj, Duary:2023gqg}. 

For massless scalar fields, the incoming, and outgoing creation/annihilation operators are given by \cite{Hijano:2019qmi, Hijano:2020szl, Li:2021snj, Duary:2023gqg}
\begin{equation}
\label{masslessfl}
\begin{split}
\sqrt{2 \omega_{\vec{p}}} \,a_{\text{in}, \vec{p}} &= c_- \int_{-\pi}^{0} d\tau \, e^{i \omega_{\vec{p}} L(\tau + \frac{\pi}{2})} \mathcal{O}^- (\tau, -\hat{p})\\
\sqrt{2 \omega_{\vec{p}}} \,a_{\text{in}, \vec{p}}^\dagger &= c_+ \int_{-\pi}^{0} d\tau \, e^{-i \omega_{\vec{p}} L(\tau + \frac{\pi}{2})} \mathcal{O}^+ (\tau, \hat{p})\\
\sqrt{2 \omega_{\vec{p}}} \,a_{\text{out}, \vec{p}} &= c_+ \int_{0}^{\pi} d\tau \, e^{i \omega_{\vec{p}} L(\tau - \frac{\pi}{2})} \mathcal{O}^- (\tau, \hat{p})\\
\sqrt{2 \omega_{\vec{p}}} \,a_{\text{out}, \vec{p}}^\dagger &= c_- \int_{0}^{\pi} d\tau \, e^{-i \omega_{\vec{p}} L(\tau - \frac{\pi}{2})} \mathcal{O}^+ (\tau, \hat{p}).
\end{split}
\end{equation}

Here, $\omega_{\vec{p}}$ is the energy of particle, $\hat{p}$ represents the direction of the momentum $\vec{p}$, while $-\hat{p}$ refers to the opposite direction, and the constant is $c_{\pm} = \pm i \frac{\sqrt{24}}{L \omega_{\vec{p}}}$. 
For massive scalar fields, the outgoing creation operator is given by \cite{Hijano:2019qmi, Hijano:2020szl, Li:2021snj, Duary:2023gqg}
\begin{equation}
\label{massivefl}
\sqrt{2 \omega_{\vec{p}}}\, \hat{a}_{\text {out }, \vec{p}}^{\dagger}=c_{-}(L, m,|\vec{p}|) \int d \tau e^{i \omega_{\vec{p}} L\left[\frac{\pi}{2}+\frac{i}{2} \log \left(\frac{\omega_{\vec{P}}+m}{\omega_{\vec{p}}-m}\right)-\tau \right]} \mathcal{O}^{+}\left(\tau, \hat{p}\right),
\end{equation}
where the constant $c_{\pm}$ is
$$
c_{ \pm}(L, m,|\vec{p}|)=\frac{1}{2 \pi}\left(\frac{m L}{\pi^3}\right)^{\frac{1}{4}}\left(\frac{2 m}{(\mp i)|\vec{p}|}\right)^{m L+\frac{1}{2}} L.
$$
In eq. \eqref{masslessfl}, note that the exponential term is given by

\[
i \times \text{frequency of the particle} \times \text{AdS radius} \times \left(\tau \pm \frac{\pi}{2}\right).
\]

It is important to note that as one takes the AdS radius to infinity (approaching the flat limit), the exponential becomes highly oscillatory due to a large phase. Consequently, the integral will be primarily influenced by values of \(\tau\) that are near \(\pm \frac{\pi}{2}\). In essence, we can interpret this formula as being governed by time intervals of order \(\mathcal{O}(1/L)\) where \(\tau \pm \frac{\pi}{2}\) is small. This indicates that global time is situated around \(\tau = \frac{\pi}{2}\) for the outgoing creation/annihilation operator (with \(\tau = -\frac{\pi}{2}\) for the incoming operator). For massless fields, in the CFT, we need to insert an operator within a window of size \(\mathcal{O}(1/L)\) at \(\tau = \pm \frac{\pi}{2}\). This applies to massless particles, and for massive particles, the situation is similar, albeit with a nuance. The limit for massive and massless particles differs slightly; the conformal dimension of a CFT operator is linked to the mass of the scalar field in AdS via

\[
m^2L^2 = \Delta(\Delta - d) \implies \Delta = \frac{d}{2} + mL + \mathcal{O}(L)^{-1}.
\]

As \(L\) approaches infinity, the conformal dimension of the CFT operator scales linearly with \(L\), indicating that the operator becomes ``heavy'' in the large AdS radius limit.

In this limit, we must consider what influences the integral. Massive particle outgoing creation operator in eq.\eqref{massivefl} shows that it will be dominated by a window of \(\mathcal{O}(1/L)\) at 

\[
\begin{split}
\text{Re}(\tau) &= \frac{\pi}{2}, \\
\text{Im}(\tau) &= \frac{1}{2} \log\left( \frac{\omega_{\vec{p}} + m}{\omega_{\vec{p}} - m} \right).
\end{split}
\]

The real part of \(\tau\) must be set to \(\frac{\pi}{2}\), while the imaginary part of \(\tau\) must also be accounted for.

Inspired by this, we propose a geometric approach to derive the formulas for creation and annihilation operators based on the geodesics of particles. We now explain the reasoning behind the formulas for the creation and annihilation operators. The time-dependent component of the exponential can be understood by calculating the time it takes for a particle to reach the boundary of AdS. This perspective provides an intuitive understanding of the flat limit of AdS. Rather than reconstructing bulk operators in AdS, we can directly extract the creation and annihilation modes in flat space. Regarding particle propagation, we show that massless particles travel along null geodesics and hit the boundary at \(\frac{\pi}{2}\). In contrast, massive particles follow timelike geodesics, which never actually reach the boundary. They oscillate indefinitely around the center of AdS. The reason they do not encounter the boundary of AdS is that their trajectories hit the boundary at complex points in the global time coordinates, represented by
\[
\tau = \frac{\pi}{2} + \frac{i}{2} \log\left(\frac{\omega_{\vec{p}} + m}{\omega_{\vec{p}} - m}\right).
\]
The geodesic hits a point in the AdS boundary where an operator needs to be inserted to create a scattering state for massive and massless particles. 

The flat limit of AdS/CFT explores the behavior of high-energy particles in AdS space, where their interactions occur at scales much smaller than the AdS radius, making the AdS curvature negligible. To model this, we construct wavepackets from narrow strips at the AdS boundary, with their distance adjusted so interactions take place in regions much smaller than the AdS curvature radius. These wavepackets travel radially along null and timelike geodesics. 

We observe a parallel between the \(\hbar \to 0\) limit in quantum mechanics and the flat limit. In the limit where \(\hbar \to 0\), the dynamics of the system increasingly reflect classical behavior, as classical paths dominate the path integral. Similarly, in the infinite radius limit, the dynamics of fields and particles are governed by classical geodesics. In the path integral formulation, contributions from paths that deviate from the classical path tend to oscillate rapidly and largely cancel each other out, leaving the classical path as the main contributor. This mirrors the behavior in the large $L$ limit, where oscillatory modes lead to a focus on specific time intervals or trajectories. Both limits emphasize the importance of geodesics. In quantum mechanics, classical paths correspond to geodesics in configuration space. In large $L$ limit, massless particles travel along null geodesics, while massive particles oscillate around the center, indicating a similar dominance of specific geometric trajectories.
The classical limit is given by 
   \[
   \hbar \to 0 \Rightarrow \text{Classical paths dominate in path integral.}
   \]
   \[
   L \to \infty \Rightarrow \text{Geodesics dominate in AdS in the flat limit.}
   \]
Considering oscillatory contributions for path integrals we have
   \[
   \int \mathcal{D}x e^{\frac{i}{\hbar} S[x]} \approx e^{\frac{i}{\hbar} S[x_{\text{cl}}]},
   \]
   where contributions from non-classical paths cancel.
Here, in the flat limit we have geodesic dominance for both massless, and massive particles. 









Let us now consider the problem of massless scattering in flat space. In both classical and quantum theory, the scattering problem is concerned with specifying initial data at \( \mathscr{I}^- \), which represents past null infinity and evolving the data to  \( \mathscr{I}^+ \) future null infinity.\footnote{In the conformal compactification of flat space null infinities $\mathscr{I^{\pm}}$ are smooth null boundaries while spatial infinity $i^0$ can be thought of as the vertex of
“the light cone at infinity”. The conformal completion of flat spacetime is smooth (in
fact, analytic) at $i^0$.} In the far past, fields and particles spread out and become widely separated, meaning that interactions are weak or negligible.\footnote{For long-range interactions in QFT and quantum gravity, the asymptotic Hamiltonian is not free. This leads to the famous IR divergences\cite{Duch:2019wpf,Prabhu:2022zcr}.} Over time, incoming particles or wavepackets move toward each other, engage in a complex sequence of interactions, and ultimately emerge at future null infinity \( \mathscr{I}^+ \). The goal in classical scattering theory is to establish a well-defined map between the phase space at \( \mathscr{I}^- \) and the phase space at \( \mathscr{I}^+ \). In quantum scattering theory, this map is described by the \( \mathcal{S}\)-matrix, which encodes the evolution from incoming states at \( \mathscr{I}^- \) to outgoing states at \( \mathscr{I}^+ \).

In this framework, the region near spatial infinity \( i^0 \) is crucial. To predict the final data on \( \mathscr{I}^+ \) based on initial data at \( \mathscr{I}^- \), we must understand how these different regions are connected. Specifically, we need to identify the matching conditions that relate the fields at \( \mathscr{I}^- \) to those at \( \mathscr{I}^+ \). A simple assumption might be that these fields are equal across \( \mathscr{I}^- \) and \( \mathscr{I}^+ \), but this turns out to be incorrect. Such a matching condition fails to respect Lorentz invariance, which is essential for consistency in relativistic theories. 

Strominger conjectured that data at the past boundary of $\mathscr{I^+}$ should be matched antipodally to data at the future of past null infinity \cite{Strominger:2013jfa}\footnote{A shockwave entering the spacetime at an angle $\hat{x}$ will leave the spacetime at the antipodal point $-\hat{x}$. Whether this matching will occur between $\mathscr{I^+_-}$ and $\mathscr{I^-_+}$ was not too obvious.}. A nice example is the Li\'enard-Wiechert field which shows how smooth fields can exhibit this antipodal matching due to the order of limits ($u \to -\infty$ or $v \to +\infty$), see fig \ref{fig:antipodalflat}. This conjecture has now been proved in both gravity and QED for a large class of spacetimes\cite{Campiglia:2017mua,Prabhu:2018gzs}\footnote{In this paper, we will be concerned only with classical Maxwell solutions on flat background.}.
Moreover, this antipodal identification of field configurations near \(i^0\) formed a crucial ingredient for the discovery of the connection between asymptotic symmetries of gauge and gravity theories and quantum soft theorems (see \cite{Strominger:2017zoo} for a review).

In this paper, we explore the antipodal matching of flat space Li\'enard-Wiechert fields from AdS geodesics. We find the Li\'enard-Wiechert solutions in both flat space and AdS using the boosting method.  This method involves identifying rest frames for a moving charge and an observer, then relating them through spacetime isometries. First, we derive the Li\'enard-Wiechert solutions in flat space due to a moving charge at the instant it passes through the origin and demonstrate that the field strength is discontinuous in large $r$ limit. The value for field strength depends upon how we approach the limit. In particular, we show that the field strengths are antipodally matched between  \( \mathscr{I}^{-}_{+} \) $(u=-\infty, \hat{x})$ and  \( \mathscr{I}^{+}_{-} \) $(v=+\infty, -\hat{x})$.  



In AdS, our approach will be to first find the solution for a static charge placed at the origin. We then use the isometries of AdS to transform this solution into one that describes a particle moving along a general timelike geodesic. We show that the Li\'enard-Wiechert fields are matched antipodally between two regions, separated by a global time difference of $\pi$. Going to the boundary of AdS along null geodesics, in the flat limit, this antipodal matching results in the flat space antipodal matching near spatial infinity.







\noindent\makebox[\linewidth]{\rule{0.9\textwidth}{1pt}} 

\subsection{Organization of the paper}
The remainder of the paper is organized as follows. In \S \ref{glads}, we discuss the geometry of AdS, global coordinates and the flat limit. 
In \S \ref{emgeo}, we study geodesics in the AdS embedding space. In \S \ref{geoglo}, we study null and timelike geodesics in global AdS and calculate the travel time from the origin to the AdS boundary. In \S \ref{cftregmap}, we study the mapping between the CFT regions and asymptotic boundary of flat space regions using the geodesics in AdS. In \S \ref{correlatorsc}, we compute the flat limit of CFT correlators using the spectral decomposition of boundary operators. In \S \ref{adsanti}, we study antipodal matching of Li\'enard-Wiechert fields in flat space and AdS using spacetime isometries and show the origins of flat space antipodal matching in AdS. Finally, in \S \ref{concl}, we discuss our conclusions and several promising future directions. 
\section{The global AdS and the flat limit}
\label{glads}
In this section, we study the global AdS and the flat limit.
 The metric for a five-dimensional Minkowski space $\mathcal{R}^{3,2}$ in Cartesian coordinates with two timelike directions is
 \begin{equation}
 ds^2=-(dX^1)^2-(dX^2)^2+(dX^3)^2+(dX^4)^2+(dX^5)^2.
 \end{equation}
 Lorentzian $\mathrm{AdS}_{4}$ is defined as an embedded ``hyperboloid'' in this flat-space given by
 \begin{equation}\label{eq:hyperboloid}
 -(X^1)^2-(X^2)^2+(X^3)^2+(X^4)^2+(X^5)^2=-L^2.
 \end{equation}
   
The constant $L$ has dimension of length and determines the curvature scale of AdS. $SO(3,2)$ is the isometry group of $\mathcal{R}^{3,2}$ and implicitly preserve the hyperboloid eq. \eqref{eq:hyperboloid}. We can define coordinates $(\rho,\tau,\Omega)$ known as global coordinates as follows
\begin{align}\label{eq:embeddingglobal}
 \begin{split}
 X_1&={L\frac{\cos\tau}{\cos\rho}}, \qquad X_{2}=L\frac{\sin\tau}{\cos\rho} \\  X_{3}&=L\tan\rho\sin\theta\cos\phi,\qquad X_{4}=L\tan\rho\sin\theta\sin\phi, \qquad X_{5}=L\tan\rho\cos\theta. 
 \end{split}
 \end{align}
In these coordinates, the line element of vacuum $\mathrm{AdS}_4$ is expressed as
\begin{equation}
\label{AdS}
ds^2=\frac{L^2}{\cos^2\rho} (-d\tau^2+d\rho^2+\sin^2\rho~ d\Omega_{2}^2).
\end{equation}
The holographic coordinate $\rho$ has the property that the boundary CFT exists at $\rho=\frac{\pi}{2}$. The coordinates used to describe the CFT are $X = ( \tau, \Omega_{2})$. It becomes apparent that the causal structure of $\mathrm{AdS}_4$ resembles that of a ``solid cylinder'' with a metric
\begin{equation}
	\label{cyl}
	ds^2=-d\tau^2+d\rho^2+\sin^2\rho~ d\Omega_{2}^2.
\end{equation}
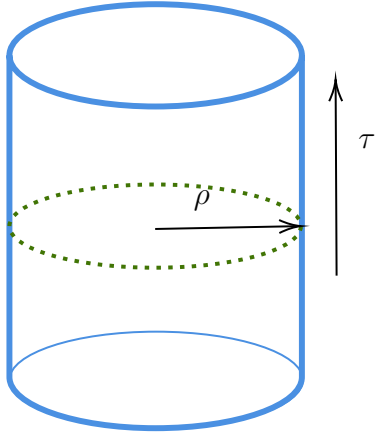
\begin{figure}[H]
	\centering
\tikzset{every picture/.style={line width=0.75pt}} 

\begin{tikzpicture}[x=0.75pt,y=0.75pt,yscale=-1,xscale=1]
	
	\draw  [color={rgb, 255:red, 74; green, 144; blue, 226 }  ,draw opacity=1 ][line width=2.25]  (393,101.81) -- (393,264.19) .. controls (393,278.44) and (359.98,290) .. (319.25,290) .. controls (278.52,290) and (245.5,278.44) .. (245.5,264.19) -- (245.5,101.81) .. controls (245.5,87.56) and (278.52,76) .. (319.25,76) .. controls (359.98,76) and (393,87.56) .. (393,101.81) .. controls (393,116.07) and (359.98,127.63) .. (319.25,127.63) .. controls (278.52,127.63) and (245.5,116.07) .. (245.5,101.81) ;
	\draw  [color={rgb, 255:red, 65; green, 117; blue, 5 }  ,draw opacity=1 ][dash pattern={on 1.69pt off 2.76pt}][line width=1.5]  (245.5,188) .. controls (245.5,176.4) and (278.41,167) .. (319,167) .. controls (359.59,167) and (392.5,176.4) .. (392.5,188) .. controls (392.5,199.6) and (359.59,209) .. (319,209) .. controls (278.41,209) and (245.5,199.6) .. (245.5,188) -- cycle ;
	\draw    (410.5,213) -- (410.01,116) ;
	\draw [shift={(410,114)}, rotate = 89.71] [color={rgb, 255:red, 0; green, 0; blue, 0 }  ][line width=0.75]    (10.93,-3.29) .. controls (6.95,-1.4) and (3.31,-0.3) .. (0,0) .. controls (3.31,0.3) and (6.95,1.4) .. (10.93,3.29)   ;
	\draw  [color={rgb, 255:red, 74; green, 144; blue, 226 }  ,draw opacity=1 ] (245.5,265.64) .. controls (245.5,252.19) and (278.63,241.28) .. (319.5,241.28) .. controls (360.37,241.28) and (393.5,252.19) .. (393.5,265.64) .. controls (393.5,279.09) and (360.37,290) .. (319.5,290) .. controls (278.63,290) and (245.5,279.09) .. (245.5,265.64) -- cycle ;
	\draw    (319,189.45) -- (390.5,188.04) ;
	\draw [shift={(392.5,188)}, rotate = 178.87] [color={rgb, 255:red, 0; green, 0; blue, 0 }  ][line width=0.75]    (10.93,-3.29) .. controls (6.95,-1.4) and (3.31,-0.3) .. (0,0) .. controls (3.31,0.3) and (6.95,1.4) .. (10.93,3.29)   ;
	
	\draw (420,140.4) node [anchor=north west][inner sep=0.75pt]    {$\tau $};
	\draw (337,168.4) node [anchor=north west][inner sep=0.75pt]    {$\rho $};

\end{tikzpicture}
	\caption{Soup-can picture of AdS: the causal structure of $\mathrm{AdS}_4$.}
	\label{fig:cylinderdiag}
\end{figure}
Fig.\ref{fig:cylinderdiag} represents the global AdS coordinate.

The benefit of utilizing global AdS is that it offers a background for the boundary CFT in the form of $\mathbb{R} \times \mathbb{S}^2$, which is 
\begin{equation}
\partial(\mathrm{AdS}_{4})=\mathbb{R} \times \mathbb{S}^2.
\end{equation}
In other words, the line element of the boundary CFT can be represented as 
\begin{equation}
ds^2_{\text{CFT}_3}=-d\tau^2+d\Omega_2^2.
\end{equation}

The interior of the cylinder represents the bulk of the $\mathrm{AdS}_4$ spacetime, whereas the cylinder itself corresponds to the boundary.

In terms of the geometry, the flat limit is achieved by taking the AdS radius to infinity, causing the line element of eq.\eqref{AdS} to resemble that of flat space. Specifically, the following transformations are applied
\begin{equation}
\tau=\frac{t}{L}~~\text{and}~~\tan\rho=\frac{r}{L}.
\end{equation}
As the AdS radius ($L$) tends to infinity, the line element transforms to
\begin{equation}
ds^2 \xrightarrow[L\to\infty]~-dt^2+dr^2+r^2d\Omega_{2}^2.
\end{equation} 
This process can be interpreted as a zooming in of the central region of AdS ($r\ll L$), wherein the geometry assumes a locally flat nature. By restricting our observations to scales minuscule in relation to the characteristic length scale of $\mathrm{AdS}_4$, we are able to get a geometric structure of $\mathcal{R}^{3,1}$.

\section{Geodesics in AdS embedding space}
\label{emgeo}
In this section, we study geodesics in AdS embedding space.
Any timelike geodesic on AdS space can be described as a curve in the embedding space $\mathcal{R}^{3,2}$. Following \cite{Sokolowski}, we find the geodesics in AdS by extremizing the Lagrangian $\frac{1}{2}\eta_{AB}\dot{X^{A}}\dot{X^{B}}$ subject to the constraint eq. \eqref{eq:hyperboloid}\footnote{$\dot{X}^A=\frac{dX^A}{ds}$, $s$ is the proper time.}
\begin{equation}
L= \frac{1}{2}\eta_{AB}\dot{X^{A}}\dot{X^{B}}-\mu\,G\left(X^{A}\right)
\end{equation}
where $G\left(X^{A}\right)=\eta_{AB}{X^{A}}{X^{B}}+L^2=0$ and $\mu$ is the Lagrange multiplier. The geodesic equations are then
\begin{equation}\label{eq:geodesic}
\ddot{X^{A}}+2\mu\,X^{A}=0.
\end{equation}
The second derivative of constraint is
\begin{equation}\label{eq:constraint}
\ddot{G}=2\left(\eta_{AB}\ddot{X^{A}}X^{B}+\eta_{AB}\dot{X^{A}}\dot{X^{B}}\right)=0.
\end{equation}

\begin{figure}[H]
	\centering
\tikzset{every picture/.style={line width=0.75pt}} 
\begin{tikzpicture}[x=0.75pt,y=0.75pt,yscale=-1,xscale=1]

\draw   (407.52,261.65) -- (407.24,386.24) -- (304.43,386.09) -- (304.71,261.5) -- cycle ;
\draw  [dash pattern={on 0.84pt off 2.51pt}]  (303,68) -- (304.71,261.51) ;
\draw  [dash pattern={on 0.84pt off 2.51pt}]  (408,51) -- (407.52,261.65) ;
\draw  [dash pattern={on 0.84pt off 2.51pt}]  (304.71,257.95) -- (407.52,258.09) ;
\draw    (275.56,275.62) -- (274.85,249.72) ;
\draw [shift={(274.8,247.72)}, rotate = 88.44] [color={rgb, 255:red, 0; green, 0; blue, 0 }  ][line width=0.75]    (10.93,-3.29) .. controls (6.95,-1.4) and (3.31,-0.3) .. (0,0) .. controls (3.31,0.3) and (6.95,1.4) .. (10.93,3.29)   ;
\draw    (340.41,398.7) -- (374.1,399.26) ;
\draw [shift={(376.1,399.29)}, rotate = 180.95] [color={rgb, 255:red, 0; green, 0; blue, 0 }  ][line width=0.75]    (10.93,-3.29) .. controls (6.95,-1.4) and (3.31,-0.3) .. (0,0) .. controls (3.31,0.3) and (6.95,1.4) .. (10.93,3.29)   ;
\draw [color={rgb, 255:red, 208; green, 2; blue, 27 }  ,draw opacity=1 ]   (304.43,386.09) -- (408,311.44) ;
\draw [color={rgb, 255:red, 208; green, 2; blue, 27 }  ,draw opacity=1 ]   (303.19,230.71) -- (408,311.44) ;
\draw [color={rgb, 255:red, 208; green, 2; blue, 27 }  ,draw opacity=1 ]   (303.19,230.71) -- (407,153) ;
\draw [color={rgb, 255:red, 208; green, 2; blue, 27 }  ,draw opacity=1 ]   (298.68,76.1) -- (407,153) ;
\draw  [color={rgb, 255:red, 189; green, 16; blue, 224 }  ,draw opacity=1 ] (306.38,385.57) .. controls (354.01,332.53) and (352.5,280.92) .. (301.86,230.74) ;
\draw  [color={rgb, 255:red, 189; green, 16; blue, 224 }  ,draw opacity=1 ] (303.19,230.71) .. controls (359.67,177.48) and (358.17,125.94) .. (298.68,76.1) ;
\draw  [dash pattern={on 0.84pt off 2.51pt}]  (302,152) -- (407,153) ;
\draw  [dash pattern={on 0.84pt off 2.51pt}]  (303,310.44) -- (408,311.44) ;
\draw    (377,132) .. controls (416.6,102.3) and (436.6,160.81) .. (475.81,132.87) ;
\draw [shift={(477,132)}, rotate = 143.13] [color={rgb, 255:red, 0; green, 0; blue, 0 }  ][line width=0.75]    (10.93,-3.29) .. controls (6.95,-1.4) and (3.31,-0.3) .. (0,0) .. controls (3.31,0.3) and (6.95,1.4) .. (10.93,3.29)   ;
\draw    (337.26,185.82) -- (466,183.04) ;
\draw [shift={(468,183)}, rotate = 178.76] [color={rgb, 255:red, 0; green, 0; blue, 0 }  ][line width=0.75]    (10.93,-3.29) .. controls (6.95,-1.4) and (3.31,-0.3) .. (0,0) .. controls (3.31,0.3) and (6.95,1.4) .. (10.93,3.29)   ;

\draw (268.32,279.36) node [anchor=north west][inner sep=0.75pt]  [font=\footnotesize]  {$\tau $};
\draw (328.52,390.57) node [anchor=north west][inner sep=0.75pt]  [font=\footnotesize,rotate=-1.54]  {$\rho $};
\draw (282,69.4) node [anchor=north west][inner sep=0.75pt]  [font=\footnotesize]  {$\pi $};
\draw (286,223.4) node [anchor=north west][inner sep=0.75pt]  [font=\footnotesize]  {$0$};
\draw (280,378.4) node [anchor=north west][inner sep=0.75pt]  [font=\footnotesize]  {$-\pi $};
\draw (279,138.4) node [anchor=north west][inner sep=0.75pt]  [font=\footnotesize]  {$\frac{\pi }{2}$};
\draw (274,299.4) node [anchor=north west][inner sep=0.75pt]  [font=\footnotesize]  {$-\frac{\pi }{2}$};
\draw (413,371.4) node [anchor=north west][inner sep=0.75pt]  [font=\footnotesize]  {$\frac{\pi }{2}$};
\draw (466,116) node [anchor=north west][inner sep=0.75pt]  [font=\footnotesize] [align=left] {Null geodesic};
\draw (471,175) node [anchor=north west][inner sep=0.75pt]  [font=\footnotesize] [align=left] {Timelike geodesic};

\end{tikzpicture}

	\caption{Timelike geodesics and null geodesics in AdS.}
	\label{fig:geodia}
\end{figure}
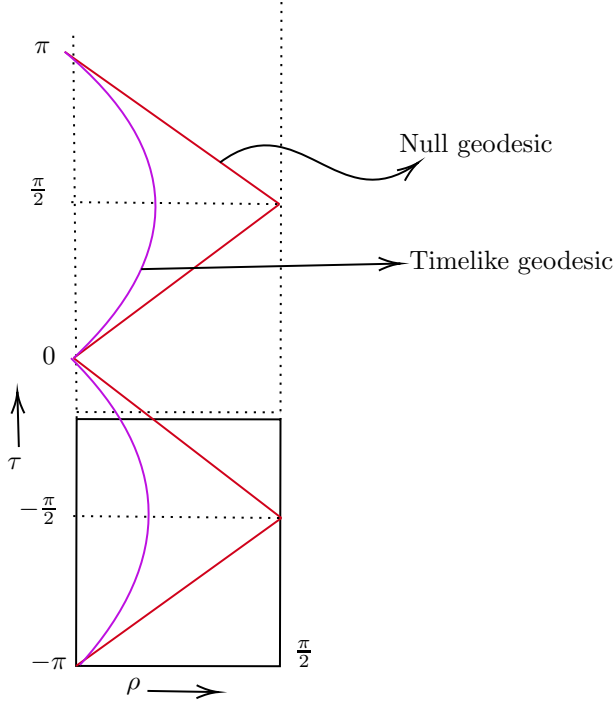
Multiplying eq.\eqref{eq:geodesic} with $\eta_{AB}X^{B}$ and using $\eta_{AB}\dot{X^{A}}\dot{X^{B}}=-1$ for timelike geodesics in eq.\eqref{eq:constraint}, we get $\mu=\frac{1}{2L^2}$ and timelike geodesic equations are
\begin{equation}\label{eq:oscillator}
\ddot{X^{A}}+\frac{1}{L^2} X^{A}=0.
\end{equation}
General Timelike geodesics are
\begin{equation}
X^{A}=a^{A}\cos\left(\frac{s}{L}\right)+b^{A}\sin\left(\frac{s}{L}\right).
\end{equation}
where $a^{A}$ and $b^{A}$ are constants satisfying the conditions
\begin{equation}
a^{A}a_{A}=b^{A}b_{A}=L^2, \quad a^{A}b_{A}=0.
\end{equation}
 At $\tau=0$, we would like to be at the center of the scattering region i.e., near $\rho=0$ , this implies $X^1=L$ and $X^i=0, \,2\leq i\leq5$. The symmetry group acts transitively on $\mathrm{AdS}_4$, meaning that any
two points are connected by a symmetry transformation, so it is a homogeneous space. In other words if the geodesic started at some other point $P_0$, there exist a non unique transformation of $SO(3,2)$ which can take the point $P$ to $P_0$. Furthermore, we can perform a rotation to make the tangent vector to the geodesic tangent to $X_2$ axis. This results in a frame in which the geodesic is restricted to the $X_1-X_2$ Euclidean plane parametrized as
\begin{equation}\label{eq:circle}
    X^1=L\cos\left(\frac{s}{L}\right), \qquad  X^2=L\sin\left(\frac{s}{L}\right), \qquad X_i=0, \qquad 3\leq i\leq 5. 
\end{equation}
It represents a circle of radius $L$ in the embedding space. In $L\rightarrow \infty$ limit, it becomes a line which is a timelike geodesic in flat space. From eq.\eqref{eq:circle}, all the embedding coordinates remain finite for a timelike geodesic. Looking at the global coordinates, the boundary of AdS is $\rho=\frac{\pi}{2}$ limit. Hence timelike geodesics never actually reach the boundary. A particle on a timelike geodesic will need infinite energy to reach boundary. Eq. \eqref{eq:oscillator} is similar to the equation for harmonic oscillator, with frequency $\omega=\frac{1}{L}$. In the next section we will see that the inability of massive particles to reach the boundary can be thought of as their trajectories hitting the boundary at complex
points in the global time coordinates.

We draw timelike geodesics and null geodesics in fig.\ref{fig:geodia}. In fig.\ref{fig:geodia}, we see that $45^{\circ}$ lines are null geodesics, timeline geodesics are bounded inside the diamond region formed by null geodesics. 



\section{Geodesics in global AdS: travel time from the origin to the AdS boundary}
\label{geoglo}
In this section, we study null and timelike geodesics in global AdS and compute their time of travel from the origin to the AdS boundary. 
The key concept behind the flat limit of AdS/CFT is to examine high-energy particles, with respect to global AdS time, that are directed into the bulk of AdS in such a way that their interactions occur on a scale much smaller than the AdS radius. 

\subsection{Intuition behind geodesic travel time}

To get the flat limit, we construct wavepackets coming from thin strips of the AdS boundary. The distance between these strips can be adjusted so that the bulk point where interactions occur in the corresponding Witten diagrams lies within a region much smaller than the AdS curvature radius \(L\). These wavepackets are highly collimated, typically with a width on the order of \(1/L\) at the boundary, meaning they have a small spatial extent relative to the global time. Because of their high energy concentration, these wavepackets travel radially inwards and outwards, following paths along both null and timelike geodesics. This motion is highly directed and focused, making the wavepackets ideal candidates for interactions in the AdS bulk, where they can collide at specific bulk points.

When studying these collisions, we focus on the travel time of the wavepackets, both for null and timelike geodesics, as they move from the origin to the AdS boundary. These travel times are crucial in determining the dynamics of the system, and in the flat limit, they should match the expected results for flat spacetime, reflecting the fact that the interactions in the bulk are effectively insensitive to the AdS curvature at high energies.

In our approach, we can avoid the detailed steps of bulk operator reconstruction and the extraction of the flat space creation and annihilation operators, the process can be simplified by directly writing down the formulas using a saddle-point approximation. We require that integrals over global time remain highly oscillatory as \( L \to \infty \), unless operators are evaluated within regions of size \( O(1/L) \) around \( \tau = \text{geodesic travel time} \). 
This oscillatory behavior constrains the function to be of the form
\begin{equation}
\label{geotratime}
 \sqrt{2 \omega_{\vec{p}}} \,a_{\text{out}/\text{in}, \vec{p}} \sim \int d\tau \, e^{i \omega_{\vec{p}} L(\tau \mp \text{geodesic travel time})}  \mathcal{O}(\tau, \pm \hat{p}).
\end{equation}


\subsection{Null geodesics}

\begin{figure}[H]
	\centering
\tikzset{every picture/.style={line width=0.75pt}} 

\begin{tikzpicture}[x=0.75pt,y=0.75pt,yscale=-1,xscale=1]

\draw  [color={rgb, 255:red, 74; green, 144; blue, 226 }  ,draw opacity=1 ] (334,55.15) -- (334,233.85) .. controls (334,248.84) and (293.48,261) .. (243.5,261) .. controls (193.52,261) and (153,248.84) .. (153,233.85) -- (153,55.15) .. controls (153,40.16) and (193.52,28) .. (243.5,28) .. controls (293.48,28) and (334,40.16) .. (334,55.15) .. controls (334,70.14) and (293.48,82.3) .. (243.5,82.3) .. controls (193.52,82.3) and (153,70.14) .. (153,55.15) ;
\draw  [color={rgb, 255:red, 74; green, 144; blue, 226 }  ,draw opacity=1 ][line width=5.25]  (153,231.46) .. controls (153,215.15) and (193.52,201.93) .. (243.5,201.93) .. controls (293.48,201.93) and (334,215.15) .. (334,231.46) .. controls (334,247.78) and (293.48,261) .. (243.5,261) .. controls (193.52,261) and (153,247.78) .. (153,231.46) -- cycle ;
\draw [color={rgb, 255:red, 74; green, 144; blue, 226 }  ,draw opacity=1 ] [dash pattern={on 0.84pt off 2.51pt}]  (243,50) -- (244,241) ;
\draw  [color={rgb, 255:red, 248; green, 231; blue, 28 }  ,draw opacity=1 ][pattern=_bixux0kyf,pattern size=6pt,pattern thickness=0.75pt,pattern radius=0pt, pattern color={rgb, 255:red, 248; green, 231; blue, 28}] (243.5,110.5) -- (278.5,145.5) -- (243.5,180.5) -- (208.5,145.5) -- cycle ;
\draw [color={rgb, 255:red, 208; green, 2; blue, 27 }  ,draw opacity=1 ]   (284,101) .. controls (284.11,103.35) and (282.98,104.59) .. (280.63,104.7) .. controls (278.28,104.81) and (277.16,106.05) .. (277.27,108.4) .. controls (277.38,110.75) and (276.25,111.98) .. (273.9,112.09) .. controls (271.55,112.2) and (270.43,113.44) .. (270.54,115.79) .. controls (270.65,118.14) and (269.52,119.38) .. (267.17,119.49) .. controls (264.82,119.6) and (263.7,120.84) .. (263.81,123.19) .. controls (263.92,125.54) and (262.79,126.77) .. (260.44,126.88) .. controls (258.09,126.99) and (256.97,128.23) .. (257.08,130.58) .. controls (257.19,132.93) and (256.06,134.17) .. (253.71,134.28) .. controls (251.36,134.39) and (250.24,135.63) .. (250.35,137.98) .. controls (250.46,140.33) and (249.33,141.57) .. (246.98,141.68) .. controls (244.63,141.79) and (243.5,143.02) .. (243.61,145.37) -- (243.5,145.5) -- (243.5,145.5) ;
\draw  [fill={rgb, 255:red, 208; green, 2; blue, 27 }  ,fill opacity=1 ] (262,117) -- (269.81,114.79) -- (268.12,122.73) ;
\draw  [color={rgb, 255:red, 74; green, 144; blue, 226 }  ,draw opacity=1 ][line width=5.25]  (153,55.15) .. controls (153,40.16) and (193.52,28) .. (243.5,28) .. controls (293.48,28) and (334,40.16) .. (334,55.15) .. controls (334,70.14) and (293.48,82.3) .. (243.5,82.3) .. controls (193.52,82.3) and (153,70.14) .. (153,55.15) -- cycle ;
\draw  [color={rgb, 255:red, 208; green, 2; blue, 27 }  ,draw opacity=1 ][fill={rgb, 255:red, 208; green, 2; blue, 27 }  ,fill opacity=1 ] (289,80) .. controls (289,78.34) and (290.34,77) .. (292,77) .. controls (293.66,77) and (295,78.34) .. (295,80) .. controls (295,81.66) and (293.66,83) .. (292,83) .. controls (290.34,83) and (289,81.66) .. (289,80) -- cycle ;
\draw [color={rgb, 255:red, 208; green, 2; blue, 27 }  ,draw opacity=1 ]   (292,94) -- (292,66) ;
\draw  [color={rgb, 255:red, 208; green, 2; blue, 27 }  ,draw opacity=1 ][fill={rgb, 255:red, 208; green, 2; blue, 27 }  ,fill opacity=1 ] (361,134) .. controls (361,132.34) and (362.34,131) .. (364,131) .. controls (365.66,131) and (367,132.34) .. (367,134) .. controls (367,135.66) and (365.66,137) .. (364,137) .. controls (362.34,137) and (361,135.66) .. (361,134) -- cycle ;
\draw [color={rgb, 255:red, 208; green, 2; blue, 27 }  ,draw opacity=1 ]   (364,148) -- (364,120) ;

\draw (345,41.4) node [anchor=north west][inner sep=0.75pt]  [font=\footnotesize]  {$\tau =\cfrac{\pi }{2}$};
\draw (344,220.4) node [anchor=north west][inner sep=0.75pt]  [font=\footnotesize]  {$\tau =-\cfrac{\pi }{2}$};
\draw (367,128) node [anchor=north west][inner sep=0.75pt]  [font=\scriptsize] [align=left] {:Massless particles hit the boundary of AdS: $\tau =\cfrac{\pi }{2}$\ };
\draw (578,123.4) node [anchor=north west][inner sep=0.75pt]  [font=\scriptsize]  {$~~~~$};

\end{tikzpicture}

	\caption{Null geodesics in AdS global coordinates: massless particles hit the boundary of AdS at $\tau =\cfrac{\pi }{2}$.}
	\label{fig:cylinderdiagnull}
\end{figure}
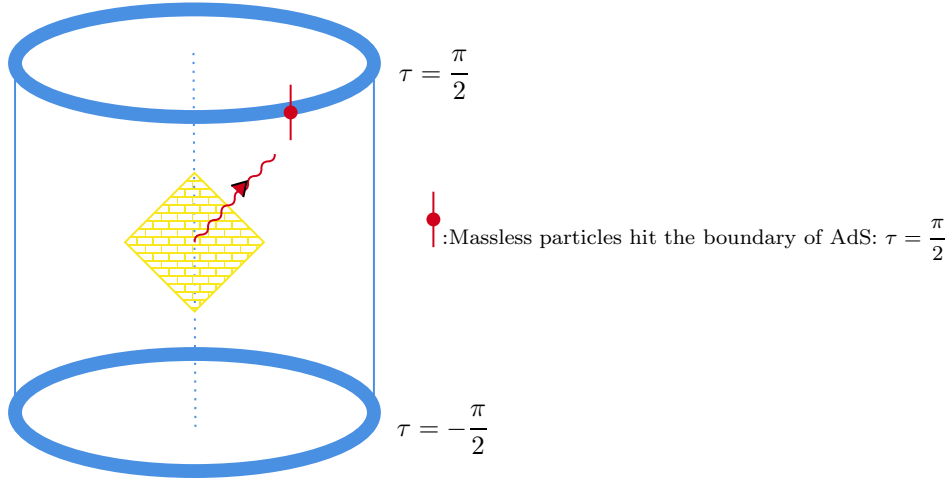

For radial null geodesics we have 
\begin{equation}
\begin{split}
ds^2&=0\\
\implies d\tau&=\pm d\rho.
\end{split}
\end{equation}
Therefore, massless particles hit the boundary of AdS at  
\begin{equation}
\Delta \tau=\pm \frac{\pi}{2}.
\end{equation}
We draw the fig. \ref{fig:cylinderdiagnull} for illustration.


\subsection{Timelike geodesics}
The line element is given by\footnote{To simplify the expressions, we set the AdS radius, $L \to 1$, since it becomes irrelevant when we express all quantities in terms of the ratio to $\frac{\omega_{\vec{p}}}{m}$.}
\begin{equation}
	ds^2=\sec^2\rho (-d\tau^2+d\rho^2+\sin^2\rho~ d\Omega_{2}^2).
\end{equation}
We consider radial motion. We have a Killing vector corresponding to the time translation 
\begin{equation}
\label{taueq}
\sec^2\rho\dot{\tau} \equiv \mathcal{E} =\frac{\omega_{\vec{p}}}{m}= \text{constant of motion}.
\end{equation}
where the dot notation represents a derivative with respect to proper time. We assume that the term $\omega_{\vec{p}}$ refers to the energy measured by an observer at rest located at $\rho=0$.

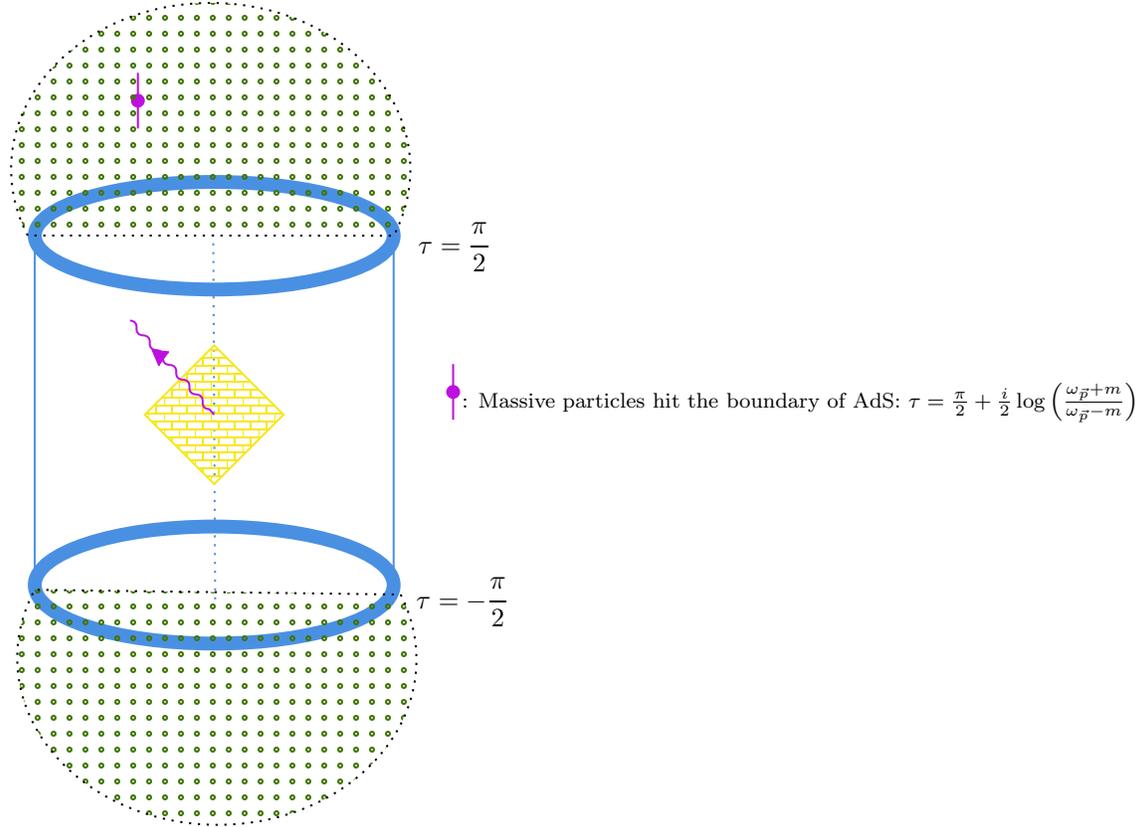
\begin{figure}[H]
	\centering
\tikzset{every picture/.style={line width=0.75pt}} 

\begin{tikzpicture}[x=0.75pt,y=0.75pt,yscale=-1,xscale=1]

\draw  [color={rgb, 255:red, 74; green, 144; blue, 226 }  ,draw opacity=1 ] (357,123.15) -- (357,301.85) .. controls (357,316.84) and (316.48,329) .. (266.5,329) .. controls (216.52,329) and (176,316.84) .. (176,301.85) -- (176,123.15) .. controls (176,108.16) and (216.52,96) .. (266.5,96) .. controls (316.48,96) and (357,108.16) .. (357,123.15) .. controls (357,138.14) and (316.48,150.3) .. (266.5,150.3) .. controls (216.52,150.3) and (176,138.14) .. (176,123.15) ;
\draw  [color={rgb, 255:red, 74; green, 144; blue, 226 }  ,draw opacity=1 ][line width=5.25]  (176,299.46) .. controls (176,283.15) and (216.52,269.93) .. (266.5,269.93) .. controls (316.48,269.93) and (357,283.15) .. (357,299.46) .. controls (357,315.78) and (316.48,329) .. (266.5,329) .. controls (216.52,329) and (176,315.78) .. (176,299.46) -- cycle ;
\draw [color={rgb, 255:red, 74; green, 144; blue, 226 }  ,draw opacity=1 ] [dash pattern={on 0.84pt off 2.51pt}]  (266,118) -- (267,309) ;
\draw  [color={rgb, 255:red, 248; green, 231; blue, 28 }  ,draw opacity=1 ][pattern=_tkc5crui7,pattern size=6pt,pattern thickness=0.75pt,pattern radius=0pt, pattern color={rgb, 255:red, 248; green, 231; blue, 28}] (266.5,178.5) -- (301.5,213.5) -- (266.5,248.5) -- (231.5,213.5) -- cycle ;
\draw [color={rgb, 255:red, 189; green, 16; blue, 224 }  ,draw opacity=1 ]   (224,166) .. controls (226.35,166.13) and (227.46,167.38) .. (227.33,169.73) .. controls (227.2,172.08) and (228.32,173.32) .. (230.67,173.45) .. controls (233.02,173.58) and (234.13,174.83) .. (234,177.18) .. controls (233.87,179.53) and (234.99,180.77) .. (237.34,180.9) .. controls (239.69,181.03) and (240.8,182.28) .. (240.67,184.63) .. controls (240.54,186.98) and (241.65,188.23) .. (244,188.36) .. controls (246.35,188.49) and (247.47,189.73) .. (247.34,192.08) .. controls (247.21,194.43) and (248.32,195.68) .. (250.67,195.81) .. controls (253.02,195.94) and (254.14,197.19) .. (254.01,199.54) .. controls (253.88,201.89) and (254.99,203.13) .. (257.34,203.26) .. controls (259.69,203.39) and (260.8,204.64) .. (260.67,206.99) .. controls (260.54,209.34) and (261.66,210.58) .. (264.01,210.71) -- (266.5,213.5) -- (266.5,213.5) ;
\draw  [color={rgb, 255:red, 189; green, 16; blue, 224 }  ,draw opacity=1 ][fill={rgb, 255:red, 189; green, 16; blue, 224 }  ,fill opacity=1 ] (238.8,188.76) -- (235.62,181.29) -- (243.71,181.96) ;
\draw  [color={rgb, 255:red, 74; green, 144; blue, 226 }  ,draw opacity=1 ][line width=5.25]  (176,123.15) .. controls (176,108.16) and (216.52,96) .. (266.5,96) .. controls (316.48,96) and (357,108.16) .. (357,123.15) .. controls (357,138.14) and (316.48,150.3) .. (266.5,150.3) .. controls (216.52,150.3) and (176,138.14) .. (176,123.15) -- cycle ;
\draw  [color={rgb, 255:red, 189; green, 16; blue, 224 }  ,draw opacity=1 ][fill={rgb, 255:red, 189; green, 16; blue, 224 }  ,fill opacity=1 ] (225,55) .. controls (225,53.34) and (226.34,52) .. (228,52) .. controls (229.66,52) and (231,53.34) .. (231,55) .. controls (231,56.66) and (229.66,58) .. (228,58) .. controls (226.34,58) and (225,56.66) .. (225,55) -- cycle ;
\draw [color={rgb, 255:red, 189; green, 16; blue, 224 }  ,draw opacity=1 ][fill={rgb, 255:red, 189; green, 16; blue, 224 }  ,fill opacity=1 ]   (228,69) -- (228,41) ;
\draw  [color={rgb, 255:red, 189; green, 16; blue, 224 }  ,draw opacity=1 ][fill={rgb, 255:red, 189; green, 16; blue, 224 }  ,fill opacity=1 ] (384,202) .. controls (384,200.34) and (385.34,199) .. (387,199) .. controls (388.66,199) and (390,200.34) .. (390,202) .. controls (390,203.66) and (388.66,205) .. (387,205) .. controls (385.34,205) and (384,203.66) .. (384,202) -- cycle ;
\draw [color={rgb, 255:red, 189; green, 16; blue, 224 }  ,draw opacity=1 ][fill={rgb, 255:red, 189; green, 16; blue, 224 }  ,fill opacity=1 ]   (387,216) -- (387,188) ;
\draw  [pattern=_7iz9i42p8,pattern size=6pt,pattern thickness=0.75pt,pattern radius=0.75pt, pattern color={rgb, 255:red, 65; green, 117; blue, 5}][dash pattern={on 0.84pt off 2.51pt}] (172.45,123.05) .. controls (167.05,112.77) and (164.06,101.41) .. (164.07,89.46) .. controls (164.09,43.15) and (209.19,5.63) .. (264.79,5.66) .. controls (320.39,5.69) and (365.44,43.26) .. (365.42,89.57) .. controls (365.41,101.51) and (362.41,112.87) .. (357,123.15) -- cycle ;
\draw  [pattern=_6ppuhd14k,pattern size=6pt,pattern thickness=0.75pt,pattern radius=0.75pt, pattern color={rgb, 255:red, 65; green, 117; blue, 5}][dash pattern={on 0.84pt off 2.51pt}] (360.53,304.27) .. controls (365.8,314.62) and (368.65,326.02) .. (368.49,337.96) .. controls (367.89,384.27) and (322.32,421.22) .. (266.73,420.49) .. controls (211.13,419.76) and (166.56,381.63) .. (167.16,335.32) .. controls (167.32,323.38) and (170.46,312.06) .. (176,301.85) -- cycle ;

\draw (368,109.4) node [anchor=north west][inner sep=0.75pt]  [font=\footnotesize]  {$\tau =\cfrac{\pi }{2}$};
\draw (367,288.4) node [anchor=north west][inner sep=0.75pt]  [font=\footnotesize]  {$\tau =-\cfrac{\pi }{2}$};
\draw (390,196) node [anchor=north west][inner sep=0.75pt]  [font=\scriptsize] [align=left] {: Massive particles hit the boundary of AdS: $\tau=\frac{\pi}{2}+\frac{i}{2}\log\Big(\frac{\omega_{\vec{p}}+m}{\omega_{\vec{p}}-m}\Big)$ };
\draw (539,210.4) node [anchor=north west][inner sep=0.75pt]  [font=\scriptsize]  {$~~$};

\end{tikzpicture}

	\caption{Timelike geodesics in AdS global coordinates: massive particles hit the boundary of AdS at $\tau =\frac{\pi}{2}+\frac{i}{2}\log\Big(\frac{\omega_{\vec{p}}+m}{\omega_{\vec{p}}-m}\Big)$.}
	\label{fig:cylinderdiagtimelike}
\end{figure}

Using the timelike geodesic condition of the four-velocity, $\dot{x}^{\mu} \dot{x}_{\mu}  = -1$, we can derive the equation governing radial motion as follows
\begin{equation}
\label{rhoeq}	
-\sec^2\rho \dot{\tau}^2+\sec^2\rho \dot{\rho}^2=-1.
\end{equation}
Using eq.\eqref{taueq} in eq.\eqref{rhoeq} we get 
\begin{equation}
\label{rhosq}
\dot{\rho}^2= \frac{\omega_{\vec{p}}^2}{m^2}\cos^4 \rho -\cos^2\rho.
\end{equation}
Using eq.\eqref{taueq} and eq.\eqref{rhosq} we get 
\begin{equation}
\frac{d\tau}{d\rho}=\pm \frac{\frac{\omega_{\vec{p}}}{m}}{\sqrt{\frac{\omega_{\vec{p}}^2}{m^2}-\sec^2\rho}}.
\end{equation}
Therefore, massive particles hit the boundary of AdS at
\begin{equation}\label{eq:complex}
\begin{split}
\Delta \tau=\pm \int_{0}^{\frac{\pi}{2}}d\rho \frac{\frac{\omega_{\vec{p}}}{m}}{\sqrt{\frac{\omega_{\vec{p}}^2}{m^2}-\sec^2\rho}}=\pm \frac{\pi}{2}\pm\frac{i}{2}\log\Big(\frac{\omega_{\vec{p}}+m}{\omega_{\vec{p}}-m}\Big).
\end{split}
\end{equation}
Thus timelike geodesics can be thought of as hitting the boundary of AdS at complex points of eq. \eqref{eq:complex}.
\paragraph{Travel time from the origin to the AdS boundary.}
The travel time for a null geodesic from the origin to the AdS boundary is 
\begin{equation}
\Delta \tau = \frac{\pi}{2}.
\end{equation}
For a timelike geodesic, the travel time from the origin to the AdS boundary is
\begin{equation}
\Delta \tau = \frac{\pi}{2} + \frac{i}{2} \log \left( \frac{\omega_{\vec{p}} + m}{\omega_{\vec{p}} - m} \right),
\end{equation}
where \(\omega_{\vec{p}}\) is the energy and \(m\) is the mass of the particle.
We draw the fig. \ref{fig:cylinderdiagtimelike} for illustration.


\section{Mapping of CFT regions with flat space regions from AdS geodesics}
\label{cftregmap}
\begin{figure}[H]
	\centering
\tikzset{every picture/.style={line width=0.75pt}} 

\begin{tikzpicture}[x=0.75pt,y=0.75pt,yscale=-1,xscale=1]

\draw  [color={rgb, 255:red, 74; green, 144; blue, 226 }  ,draw opacity=1 ] (377,143.15) -- (377,321.85) .. controls (377,336.84) and (336.48,349) .. (286.5,349) .. controls (236.52,349) and (196,336.84) .. (196,321.85) -- (196,143.15) .. controls (196,128.16) and (236.52,116) .. (286.5,116) .. controls (336.48,116) and (377,128.16) .. (377,143.15) .. controls (377,158.14) and (336.48,170.3) .. (286.5,170.3) .. controls (236.52,170.3) and (196,158.14) .. (196,143.15) ;
\draw  [color={rgb, 255:red, 74; green, 144; blue, 226 }  ,draw opacity=1 ][line width=5.25]  (196,319.46) .. controls (196,303.15) and (236.52,289.93) .. (286.5,289.93) .. controls (336.48,289.93) and (377,303.15) .. (377,319.46) .. controls (377,335.78) and (336.48,349) .. (286.5,349) .. controls (236.52,349) and (196,335.78) .. (196,319.46) -- cycle ;
\draw [color={rgb, 255:red, 74; green, 144; blue, 226 }  ,draw opacity=1 ] [dash pattern={on 0.84pt off 2.51pt}]  (286,138) -- (287,329) ;
\draw  [color={rgb, 255:red, 248; green, 231; blue, 28 }  ,draw opacity=1 ][pattern=_icy489xm3,pattern size=6pt,pattern thickness=0.75pt,pattern radius=0pt, pattern color={rgb, 255:red, 248; green, 231; blue, 28}] (286.5,198.5) -- (321.5,233.5) -- (286.5,268.5) -- (251.5,233.5) -- cycle ;
\draw [color={rgb, 255:red, 189; green, 16; blue, 224 }  ,draw opacity=1 ]   (244,186) .. controls (246.35,186.13) and (247.46,187.38) .. (247.33,189.73) .. controls (247.2,192.08) and (248.32,193.32) .. (250.67,193.45) .. controls (253.02,193.58) and (254.13,194.83) .. (254,197.18) .. controls (253.87,199.53) and (254.99,200.77) .. (257.34,200.9) .. controls (259.69,201.03) and (260.8,202.28) .. (260.67,204.63) .. controls (260.54,206.98) and (261.65,208.23) .. (264,208.36) .. controls (266.35,208.49) and (267.47,209.73) .. (267.34,212.08) .. controls (267.21,214.43) and (268.32,215.68) .. (270.67,215.81) .. controls (273.02,215.94) and (274.14,217.19) .. (274.01,219.54) .. controls (273.88,221.89) and (274.99,223.13) .. (277.34,223.26) .. controls (279.69,223.39) and (280.8,224.64) .. (280.67,226.99) .. controls (280.54,229.34) and (281.66,230.58) .. (284.01,230.71) -- (286.5,233.5) -- (286.5,233.5) ;
\draw  [color={rgb, 255:red, 189; green, 16; blue, 224 }  ,draw opacity=1 ][fill={rgb, 255:red, 189; green, 16; blue, 224 }  ,fill opacity=1 ] (258.8,208.76) -- (255.62,201.29) -- (263.71,201.96) ;
\draw  [color={rgb, 255:red, 74; green, 144; blue, 226 }  ,draw opacity=1 ][line width=5.25]  (196,143.15) .. controls (196,128.16) and (236.52,116) .. (286.5,116) .. controls (336.48,116) and (377,128.16) .. (377,143.15) .. controls (377,158.14) and (336.48,170.3) .. (286.5,170.3) .. controls (236.52,170.3) and (196,158.14) .. (196,143.15) -- cycle ;
\draw  [color={rgb, 255:red, 189; green, 16; blue, 224 }  ,draw opacity=1 ][fill={rgb, 255:red, 189; green, 16; blue, 224 }  ,fill opacity=1 ] (245,75) .. controls (245,73.34) and (246.34,72) .. (248,72) .. controls (249.66,72) and (251,73.34) .. (251,75) .. controls (251,76.66) and (249.66,78) .. (248,78) .. controls (246.34,78) and (245,76.66) .. (245,75) -- cycle ;
\draw [color={rgb, 255:red, 189; green, 16; blue, 224 }  ,draw opacity=1 ][fill={rgb, 255:red, 189; green, 16; blue, 224 }  ,fill opacity=1 ]   (248,89) -- (248,61) ;
\draw  [color={rgb, 255:red, 189; green, 16; blue, 224 }  ,draw opacity=1 ][fill={rgb, 255:red, 189; green, 16; blue, 224 }  ,fill opacity=1 ] (15,235) .. controls (15,233.34) and (16.34,232) .. (18,232) .. controls (19.66,232) and (21,233.34) .. (21,235) .. controls (21,236.66) and (19.66,238) .. (18,238) .. controls (16.34,238) and (15,236.66) .. (15,235) -- cycle ;
\draw [color={rgb, 255:red, 189; green, 16; blue, 224 }  ,draw opacity=1 ][fill={rgb, 255:red, 189; green, 16; blue, 224 }  ,fill opacity=1 ]   (18,249) -- (18,221) ;
\draw  [pattern=_ar6uez0vk,pattern size=6pt,pattern thickness=0.75pt,pattern radius=0.75pt, pattern color={rgb, 255:red, 65; green, 117; blue, 5}][dash pattern={on 0.84pt off 2.51pt}] (192.45,143.05) .. controls (187.05,132.77) and (184.06,121.41) .. (184.07,109.46) .. controls (184.09,63.15) and (229.19,25.63) .. (284.79,25.66) .. controls (340.39,25.69) and (385.44,63.26) .. (385.42,109.57) .. controls (385.41,121.51) and (382.41,132.87) .. (377,143.15) -- cycle ;
\draw  [pattern=_bn8zr1qen,pattern size=6pt,pattern thickness=0.75pt,pattern radius=0.75pt, pattern color={rgb, 255:red, 65; green, 117; blue, 5}][dash pattern={on 0.84pt off 2.51pt}] (380.53,324.27) .. controls (385.8,334.62) and (388.65,346.02) .. (388.49,357.96) .. controls (387.89,404.27) and (342.32,441.22) .. (286.73,440.49) .. controls (231.13,439.76) and (186.56,401.63) .. (187.16,355.32) .. controls (187.32,343.38) and (190.46,332.06) .. (196,321.85) -- cycle ;
\draw [color={rgb, 255:red, 208; green, 2; blue, 27 }  ,draw opacity=1 ]   (327,189) .. controls (327.11,191.35) and (325.98,192.59) .. (323.63,192.7) .. controls (321.28,192.81) and (320.16,194.05) .. (320.27,196.4) .. controls (320.38,198.75) and (319.25,199.98) .. (316.9,200.09) .. controls (314.55,200.2) and (313.43,201.44) .. (313.54,203.79) .. controls (313.65,206.14) and (312.52,207.38) .. (310.17,207.49) .. controls (307.82,207.6) and (306.7,208.84) .. (306.81,211.19) .. controls (306.92,213.54) and (305.79,214.77) .. (303.44,214.88) .. controls (301.09,214.99) and (299.97,216.23) .. (300.08,218.58) .. controls (300.19,220.93) and (299.06,222.17) .. (296.71,222.28) .. controls (294.36,222.39) and (293.24,223.63) .. (293.35,225.98) .. controls (293.46,228.33) and (292.33,229.57) .. (289.98,229.68) .. controls (287.63,229.79) and (286.5,231.02) .. (286.61,233.37) -- (286.5,233.5) -- (286.5,233.5) ;
\draw  [color={rgb, 255:red, 208; green, 2; blue, 27 }  ,draw opacity=1 ][fill={rgb, 255:red, 208; green, 2; blue, 27 }  ,fill opacity=1 ] (309,202) -- (316.81,199.79) -- (315.12,207.73) ;
\draw  [color={rgb, 255:red, 208; green, 2; blue, 27 }  ,draw opacity=1 ][fill={rgb, 255:red, 208; green, 2; blue, 27 }  ,fill opacity=1 ] (337,165) .. controls (337,163.34) and (338.34,162) .. (340,162) .. controls (341.66,162) and (343,163.34) .. (343,165) .. controls (343,166.66) and (341.66,168) .. (340,168) .. controls (338.34,168) and (337,166.66) .. (337,165) -- cycle ;
\draw [color={rgb, 255:red, 208; green, 2; blue, 27 }  ,draw opacity=1 ]   (340,182) -- (340,154) ;
\draw  [color={rgb, 255:red, 208; green, 2; blue, 27 }  ,draw opacity=1 ][fill={rgb, 255:red, 208; green, 2; blue, 27 }  ,fill opacity=1 ] (15,199) .. controls (15,197.34) and (16.34,196) .. (18,196) .. controls (19.66,196) and (21,197.34) .. (21,199) .. controls (21,200.66) and (19.66,202) .. (18,202) .. controls (16.34,202) and (15,200.66) .. (15,199) -- cycle ;
\draw [color={rgb, 255:red, 208; green, 2; blue, 27 }  ,draw opacity=1 ]   (18,210) -- (18,182) ;
\draw  [color={rgb, 255:red, 248; green, 231; blue, 28 }  ,draw opacity=1 ][pattern=_3une2mc6d,pattern size=6pt,pattern thickness=0.75pt,pattern radius=0pt, pattern color={rgb, 255:red, 248; green, 231; blue, 28}] (541.85,174.3) -- (624.08,252) -- (541.85,329.7) -- (459.62,252) -- cycle ;
\draw [color={rgb, 255:red, 208; green, 2; blue, 27 }  ,draw opacity=1 ]   (581,212) .. controls (581.03,214.36) and (579.86,215.55) .. (577.5,215.57) .. controls (575.14,215.6) and (573.98,216.79) .. (574.01,219.15) .. controls (574.04,221.51) and (572.87,222.7) .. (570.51,222.72) .. controls (568.15,222.74) and (566.98,223.93) .. (567.01,226.29) .. controls (567.04,228.65) and (565.87,229.84) .. (563.51,229.87) .. controls (561.16,229.9) and (559.99,231.09) .. (560.02,233.44) .. controls (560.05,235.8) and (558.88,236.99) .. (556.52,237.01) .. controls (554.16,237.04) and (552.99,238.23) .. (553.02,240.59) .. controls (553.05,242.95) and (551.88,244.14) .. (549.52,244.16) .. controls (547.17,244.19) and (546,245.38) .. (546.03,247.73) .. controls (546.06,250.09) and (544.89,251.28) .. (542.53,251.31) -- (541.85,252) -- (541.85,252) ;
\draw [color={rgb, 255:red, 189; green, 16; blue, 224 }  ,draw opacity=1 ]   (503,213) .. controls (505.36,213) and (506.54,214.18) .. (506.53,216.54) .. controls (506.52,218.9) and (507.7,220.08) .. (510.06,220.08) .. controls (512.42,220.09) and (513.6,221.27) .. (513.59,223.63) .. controls (513.58,225.98) and (514.76,227.16) .. (517.11,227.17) .. controls (519.47,227.17) and (520.65,228.35) .. (520.64,230.71) .. controls (520.63,233.07) and (521.81,234.25) .. (524.17,234.25) .. controls (526.53,234.26) and (527.71,235.44) .. (527.7,237.8) .. controls (527.69,240.16) and (528.87,241.34) .. (531.23,241.34) .. controls (533.59,241.34) and (534.77,242.52) .. (534.76,244.88) .. controls (534.75,247.24) and (535.93,248.42) .. (538.29,248.42) .. controls (540.65,248.43) and (541.83,249.61) .. (541.82,251.97) -- (541.85,252) -- (541.85,252) ;
\draw  [color={rgb, 255:red, 208; green, 2; blue, 27 }  ,draw opacity=1 ][fill={rgb, 255:red, 208; green, 2; blue, 27 }  ,fill opacity=1 ] (558,228) -- (565.81,225.79) -- (564.12,233.73) ;
\draw  [color={rgb, 255:red, 189; green, 16; blue, 224 }  ,draw opacity=1 ][fill={rgb, 255:red, 189; green, 16; blue, 224 }  ,fill opacity=1 ] (518.8,235.76) -- (515.62,228.29) -- (523.71,228.96) ;
\draw  [fill={rgb, 255:red, 80; green, 227; blue, 194 }  ,fill opacity=0.23 ] (387,236) -- (429,236) -- (429,226) -- (457,246) -- (429,266) -- (429,256) -- (387,256) -- (397,246) -- cycle ;
\draw  [fill={rgb, 255:red, 80; green, 227; blue, 194 }  ,fill opacity=0.18 ] (333,244) -- (333,202) .. controls (333,192.06) and (341.06,184) .. (351,184) -- (387,184) .. controls (396.94,184) and (405,192.06) .. (405,202) -- (405,214) -- (411,214) -- (396,232) -- (381,214) -- (387,214) -- (387,202) .. controls (387,202) and (387,202) .. (387,202) -- (351,202) .. controls (351,202) and (351,202) .. (351,202) -- (351,244) -- cycle ;

\draw (388,129.4) node [anchor=north west][inner sep=0.75pt]  [font=\footnotesize]  {$\tau =\cfrac{\pi }{2}$};
\draw (387,308.4) node [anchor=north west][inner sep=0.75pt]  [font=\footnotesize]  {$\tau =-\cfrac{\pi }{2}$};
\draw (24,219.4) node [anchor=north west][inner sep=0.75pt]  [font=\scriptsize]  {$:\ \tau =\cfrac{\pi }{2} +\frac{i}{2} \ \log\left(\frac{\omega _{\vec{p}} +m}{\omega _{\vec{p}} -m}\right)$};
\draw (24,194) node [anchor=north west][inner sep=0.75pt]  [font=\scriptsize] [align=left] {: \ };
\draw (32,188.4) node [anchor=north west][inner sep=0.75pt]  [font=\scriptsize]  {$\tau =\cfrac{\pi }{2}$};
\draw (269,4.4) node [anchor=north west][inner sep=0.75pt]  [font=\footnotesize]  {$\partial \mathcal{M}_{+}$};
\draw (276,451.4) node [anchor=north west][inner sep=0.75pt]  [font=\footnotesize]  {$\partial \mathcal{M}_{-}$};
\draw (164,134.4) node [anchor=north west][inner sep=0.75pt]    {$\mathscr{I}^{+}$};
\draw (160,307.4) node [anchor=north west][inner sep=0.75pt]    {$\mathscr{I}^{-}$};
\draw (533,159.4) node [anchor=north west][inner sep=0.75pt]    {$i^{+}$};
\draw (537,333.4) node [anchor=north west][inner sep=0.75pt]    {$i^{-}$};
\draw (626,244.4) node [anchor=north west][inner sep=0.75pt]    {$i^{0}$};
\draw (585,191.4) node [anchor=north west][inner sep=0.75pt]    {$\mathscr{I}^{+}$};
\draw (586,285.4) node [anchor=north west][inner sep=0.75pt]    {$\mathscr{I}^{-}$};

\end{tikzpicture}

	\caption{Mapping of CFT regions to flat space regions. Here, \(\partial \mathcal{M}_+\) denotes future timelike infinity, and \(\partial \mathcal{M}_-\) denotes past timelike infinity. The blue fringe regions represent future and past null infinity. These regions are described by \(\tau = \pm \frac{\pi}{2}\), corresponding to future null infinity \(\mathscr{I}^+\) and past null infinity \(\mathscr{I}^-\), collectively referred to as \(\mathscr{I}^{\pm}\) on the AdS boundary. The pill-shaped regions represent analytic continuations of the boundary CFT in the imaginary direction of global time, acting as future and past timeline infinity \(i^{\pm}\).
}
	\label{fig:pillshaped}
\end{figure}
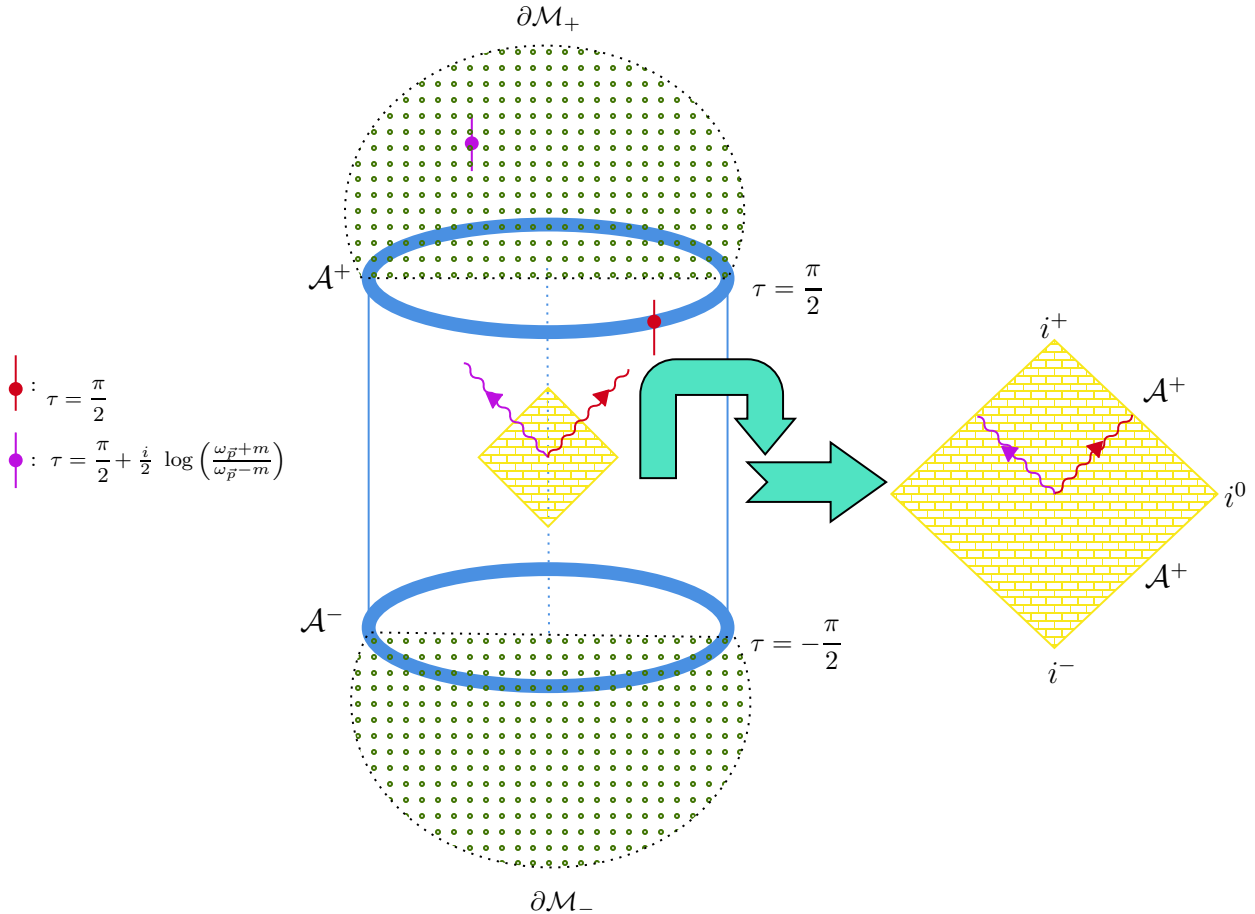

In this section, we study the mapping of CFT regions with flat space regions from AdS geodesics.
 The geodesic analysis suggests that, when considering a flat limit of AdS/CFT, there are two key regions in the CFT. One region is around $\frac{\pi}{2}$ and $-\frac{\pi}{2}$, which acts as null infinity, while the other consists of Euclidean domes that represent future and past timeline infinity in the flat limit. Regarding particle propagation, as proved earlier, massless particles move along null geodesics and reach $\tau=\frac{\pi}{2}$, while massive particles follow timelike geodesics and never actually reach the boundary-meaning they reach the boundary at a complex $\tau$-point. See fig. \ref{fig:pillshaped}.

In these fringes about $\tau=\pm\frac{\pi}{2}$, we can replace
\beq
\tau=\pm\frac{\pi}{2}+\frac{u}{L}.
\eeq
Then the boundary metric becomes
\beq
ds^2=-du^2+L^2\,d\Omega_2^2.
\eeq
In the flat limit, the metric resembles the metric of $\mathscr{I^{\pm}}$, where $u$ plays the role of retarded and advanced time\footnote{One can always locally identify two manifolds of the same dimension. But that is not what is happening here.}.

\section{Flat limit of CFT correlators}
\label{correlatorsc}
 In this section, we explore the flat limit of CFT correlators for massless and massive scattering.

\subsection{Massless scattering}
The scattering amplitude is given by the inner product
\begin{equation}\label{eq:smatrix1}
    \mathcal{G}(p_1,p_2):=\langle \phi_{\text{out}}(\mathbf{p_2})|\phi_{\text{in}}(\mathbf{p_1})\rangle.
\end{equation}
where $\phi_{\text{out}}$ and $\phi_{\text{in}}$ are the fields in far future and far past. From a normalization point of view, the above inner product can also be written as
\begin{equation}
    \mathcal{G}(p_1,p_2):=\langle \phi_{\text{in}}(\mathbf{-p_2})^{\ast}|\phi_{\text{in}}(\mathbf{p_1})\rangle.
\end{equation}
After mode decomposition, eq. \eqref{eq:smatrix1} gives
\begin{equation}
     \mathcal{G}(p_1,p_2)=\langle a_{\text{out},p_2}|a_{\text{in},p_1}^{\dag}\rangle.
\end{equation}
We can now compute scattering amplitudes from CFT correlators using eq.\eqref{geotratime} 
\begin{equation}
a_{\text{out},p_2} = C \frac{1}{\sqrt{2\omega_{\Vec{p}_2}}}\int_{0}^{\pi} d\tau e^{i\omega_{\Vec{p}_2}L(\tau-\frac{\pi}{2})}O^-(\tau,\hat{p_2}),
\end{equation}
\begin{equation}
a^\dagger_{\text{in},p_1} = C^* \frac{1}{\sqrt{2\omega_{\Vec{p}_1}}}\int_{-\pi}^{0} d\tau' e^{-i\omega_{\Vec{p}_1}L(\frac{\pi}{2}+\tau')}O^+(\tau',-\hat{p}_2).
\end{equation}

This gives 
\begin{equation}
    \mathcal{G}(p_1,p_2)= \frac{|C|^2}{\sqrt{2\omega_{\Vec{p}_1}}\sqrt{2\omega_{\Vec{p}_2}}}\int_{0}^{\pi} d\tau e^{i\omega_{\Vec{p}_2}L(\tau-\frac{\pi}{2})}\int_{-\pi}^{0} d\tau' e^{-i\omega_{\Vec{p}_1}L(\frac{\pi}{2}+\tau')}\langle O^-(\tau,\hat{p_2}) O^+(\tau',-\hat{p}_1)\rangle.
\end{equation}
Substituting $\tau-\frac{\pi}{2}=\frac{t}{L}$ and $\tau'+\frac{\pi}{2}=\frac{t'}{L}$ we get
\begin{equation}\label{eq:smatrix2}
    \mathcal{G}(p_1,p_2)= \frac{|C|^2}{\sqrt{2\omega_{\Vec{p}_1}}\sqrt{2\omega_{\Vec{p}_2}}L^2}\int_{\frac{-\pi L}{2}}^{\frac{\pi L}{2}} dt\, e^{i\omega_{\Vec{p}_2}t}\int_{\frac{-\pi L}{2}}^{\frac{\pi L}{2}}dt'\, e^{-i\omega_{\Vec{p}_1}t'}\left\langle O^-\left(\frac{t}{L}+\frac{\pi}{2},\hat{p_1}\right) O^+\left(\frac{t'}{L}-\frac{\pi}{2},-\hat{p}_2\right)\right\rangle.
\end{equation}
From spectral decomposition of operators, we have
\begin{equation}
    \left\langle O^-\left(\frac{t}{L}+\frac{\pi}{2},\hat{p_1}\right), O^+\left(\frac{t'}{L}-\frac{\pi}{2},-\hat{p}_2\right)\right\rangle=\sum_{n\in\mathbb{Z}^+,l,m}C_{n,l}\,e^{i\omega_{n,l}(\frac{t'-t}{L}-\pi)}Y^m_l(\hat{p}_1)Y^{m*}_l(-\hat{p}_2).
\end{equation}
This form is more convenient to do calculations, given that the integrals appearing in the $2$-point function are Fourier transforms with respect to global time.
Substituting back into eq. \eqref{eq:smatrix2} we get
\begin{equation}
\begin{split}
 \mathcal{G}(p_1,p_2)&= \frac{|C|^2}{\sqrt{2\omega_{\Vec{p}_1}}\sqrt{2\omega_{\Vec{p}_2}}L^2}\sum_{n\in\mathbb{Z}^+,l,m}C_{n,l}\int_{\frac{-\pi L}{2}}^{\frac{\pi L}{2}} dt\, dt'\,e^{i\left(\omega_{\Vec{p}_2}-\frac{\omega_{n,l}}{L}\right)t}\, e^{-i\left(\omega_{\Vec{p}_1}-\frac{\omega_{n,l}}{L}\right)t'} \\
 &~~~~~~~~~~~~~~~~~~~~~~~~~~~~~~~~~~e^{-i\omega_{n,l}\pi}Y^m_l(\hat{p}_1)Y^{m*}_l(-\hat{p}_2),
\end{split}
\end{equation}
Finally, we can write
\begin{equation}
\mathcal{G}(p_1,p_2) = |C|^2 (2\pi)^2 L^4 \frac{\omega_{\vec{p_1}}^4}{12} \,2\pi\delta^{(3)}(\vec{p}_1-\vec{p}_2). 
\end{equation}
The constants multiplying the delta function can be gotten rid of by normalizing the creation/annihilation operators accordingly. Indeed, choosing
\begin{equation}
C = \sqrt{\frac{12}{2\pi(L\omega_{\vec{p}_1})^2}}, 
\end{equation}
we get 
\begin{equation}
\mathcal{G}(p_1,p_2) = 2\omega_{\vec{p}_1}\delta^{(3)}(\vec{p}_1-\vec{p}_2). 
\end{equation}

\subsection{Massive scattering}
We take flat limit of CFT correlators using eq.\eqref{geotratime} 
\begin{equation}
a_{\text{out},p_2} = C \frac{1}{\sqrt{2\omega_{\vec{p}_2}}}\int_{\Gamma} d\tau e^{-i\omega_{\vec{p}_2}L\left(\frac{\pi}{2}-\frac{i}{2}\log\left(\frac{\omega_{\vec{p}_2}+m}{\omega_{\vec{p}_2}-m}\right)-\tau\right)}O^-(\tau,\hat{p}_2),
\end{equation}

\begin{equation}
a^\dagger_{\text{in},p_1} = C^* \frac{1}{\sqrt{2\omega_{\vec{p}_1}}}\int_{\Gamma'} d\tau' e^{i\omega_{\vec{p}_1}L\left(-\frac{\pi}{2}+\frac{i}{2}\log\left(\frac{\omega_{\vec{p}_1}+m}{\omega_{\vec{p}_1}-m}\right)-\tau'\right)}O^+(\tau',-\hat{p}_2).
\end{equation}
Depending on whether particles are incoming or outgoing, the insertion points are located at \( \tau = \pm \frac{\pi}{2} +  i \tilde{\tau} \), where
\begin{equation}
\tilde{\tau} = 
\begin{cases}
\frac{1}{2}\log\Big(\frac{\omega_{\vec{p}} + m}{\omega_{\vec{p}} - m}\Big), & \text{for outgoing modes} \\
-\frac{1}{2}\log\Big(\frac{\omega_{\vec{p}} + m}{\omega_{\vec{p}} - m}\Big), & \text{for incoming modes}.
\end{cases}
\end{equation}

Here, \(\tilde{\tau}\) represents a coordinate in the Euclidean half-sphere. To maintain analyticity in the scattering amplitudes, it is useful to shift the \(\tau\)-contour into the complex plane by \(\pm \frac{\pi}{2} + i \tilde{\tau}\). Specifically, for outgoing modes with positive \(\omega_{\vec{p}}\), the shift becomes
\begin{equation}
\tilde{\tau} = \frac{1}{2}\log\Big(\frac{\omega_{\vec{p}} + m}{\omega_{\vec{p}} - m}\Big),
\end{equation}
and for incoming modes with negative \(\omega_{\vec{p}}\), the shift is
\begin{equation}
\tilde{\tau} = -\frac{1}{2}\log\Big(\frac{\omega_{\vec{p}} + m}{\omega_{\vec{p}} - m}\Big).
\end{equation}
We have
\begin{equation}
     \mathcal{G}(p_1,p_2)=\langle a_{\text{out},p_2}|a_{\text{in},p_1}^{\dag}\rangle.
\end{equation}
As before we will get the two delta functions along with the factor $e^{-i\omega_{n,l}\pi}$. In massive case, we get an extra exponential term
$$e^{\frac{-\omega_{n,l}}{2}\log\left[\left(\frac{\omega_{\vec{p}_1}+m}{\omega_{\vec{p}_1}-m}\right)\left(\frac{\omega_{\vec{p}_2}+m}{\omega_{\vec{p}_2}-m}\right)\right]}$$
 Under the matching $\omega_{\vec{p}_1}=\omega_{\vec{p}_2}=L\omega_{n,l}$ due to the presence of delta functions, this term vanishes in the large $L$ limit. But if we allow $\tau$ to be imaginary and in particular if we have
\begin{equation}
\text{Im}(\tau)=\frac{1}{2}\log\left(\frac{\omega_{\vec{p}}+m}{\omega_{\vec{p}}-m}\right),
\end{equation}
for both global times $\tau$ and $\tau'$ then this term drops out and we get
\begin{equation}
     \mathcal{G}(p_1,p_2)=2\omega_{\vec{p}_1}\delta^{(3)}(\vec{p}_1-\vec{p}_2).
\end{equation}



\section{AdS origin of flat space antipodal matching for  Li\'enard-Wiechert fields}
\label{adsanti}
In this section, we study Li\'enard-Wiechert solutions in the flat limit of AdS. We compute the field strength of a moving charge by boosting a static charge using AdS isometries. First, we derive these solutions in flat spacetime, using the method of boosting a static charge to characterize the electromagnetic field of a moving point charge in flat space. After establishing this, we generalize the approach to AdS space by incorporating its isometries. In AdS, we find the solution for a static charge at the origin. Using AdS isometries, we transform this static solution into one representing a particle moving along a general timelike geodesic. 

The Li\'enard-Wiechert potentials describe the electromagnetic potentials associated with a charge moving along a general trajectory. Determining the electromagnetic field produced by a charge moving along an arbitrary path is a common problem in electrodynamics textbooks. Typically, the field is calculated by solving the wave equation to obtain the Li\'enard-Wiechert potentials, followed by computing the necessary derivatives with respect to position and time. Most textbooks derive the Li\'enard-Wiechert potentials based on the original work of Li\'enard \cite{Lienard1898} and Wiechert \cite{Wiechert1900}, see e.g., the textbooks of J. Schwinger et al. \cite{Schwinger2018}, Panofsky-Phillips \cite{Panofsky1975}, and Jackson \cite{Jackson1999}. The Li\'enard-Wiechert potentials are expressed as functions of the retarded time, which depends on both the location where the field is being evaluated and the trajectory of the charge. In this section, we present an alternative derivation of the electromagnetic field for a charge moving arbitrarily, using a manifestly Lorentz-invariant approach. The idea is that the electromagnetic fields of a charge can be derived in a Lorentz frame where the charge is momentarily at rest at the retarded time. The fields in the original frame can be obtained by performing a Lorentz transformation. This method of deriving the Li\'enard-Wiechert potentials through Lorentz transformations follows the hints first given by Minkowski \cite{Minkowski}, see also \cite{Padmanabhan:2008gr}, and \S \texttt{23.2.3 A Covariant Derivation} of Zangwill \cite{Zangwill2013}. Minkowski said that the Li\'enard-Wiechert potentials gave ``\textit{perhaps the most striking example}'' of the advantages given by Lorentz transformation, see \cite{Minkowski:1952space}. 

\subsection{Li\'enard-Wiechert solution in flat space}

To obtain the Li\'enard–Wiechert solution for a single charged particle moving with constant four velocity, we start by considering the electrostatic potential of a static charge located at the origin. For a static point particle with charge $Q$ at origin, the charge density is $Q \,\delta^{(3)}(\Vec{x})$. In this case, the Maxwell equations can be easily solved assuming spherical symmetry to give the scalar potential\footnote{We work in natural units.}

\beq
\phi= \frac{Q}{r},
\eeq
and the electric field 
\beq
\Vec{E}=-\nabla\phi=\frac{Q\hat{x}}{r^2}=\frac{Q\Vec{x}}{r^3},
\eeq
where
\beq r^2=\Vec{x}\cdot \Vec{x}, \qquad \Vec{x}=r\,\hat{x}.
\eeq

Now we can compute the vector potential for a relativistic charge at the instant when it passes through the origin by applying Lorentz transformations. We define the two inertial frames as follows:
\begin{itemize}
\item S': Rest frame of the particle at the origin with coordinates $(t', x', y', z')$\footnote{Since we will be working with moving charge, we want to keep the unprimed coordinates for the boosted frame.}.
\item S: Rest frame of the observer
\end{itemize}
In the frame S', the charged particle will have a velocity $\Vec{\beta}$ in an arbitrary direction. In other words, it means that frame S is moving with velocity $-\Vec{\beta}$ with respect to frame S'.
In the rest frame S' of the particle, the components of 4-potential are
  \beq
   \phi^{\rq}=\frac{Q}{r^{\rq}}, \qquad \Vec{A^{\rq}}=0.
  \eeq
  General Lorentz transformations going from S' to S are
\begin{eqnarray}
\begin{split}
\phi&=\gamma\left(\phi^{\rq}+\Vec{\beta} \cdot \Vec{A^{\rq}}\right) \\
\Vec{A}&=\Vec{A^{\rq}}+ \frac{\gamma-1}{\beta^2}\left(\Vec{\beta} \cdot \Vec{A^{\rq}}\right)\Vec{\beta}+\gamma\Vec{\beta}\phi^{\rq}.
\end{split}
 \end{eqnarray}
  Then $\phi$ and $\Vec{A}$ in frame S are given by
  \beq
\phi=\gamma\phi^{\rq}=\frac{\gamma Q}{r^{\rq}},\quad \Vec{A}=\frac{\gamma Q\Vec{\beta}}{r^{\rq}}.
  \eeq
  The field strength in frame S' is
\beq
 F_{r^{\rq}t^{\rq}}= \partial_{r^{\rq}}A_{t^{\rq}}-\partial_{t^{\rq}}A_{r^{\rq}}= -\partial_{r^{\rq}}\phi= \frac{Q}{r^{{\rq}^2}}=\frac{Q\Vec{x}'}{r'^3},
\eeq
which is simply the radial component of electric field $\Vec{E}$. The field strength of the boosted charge can be obtained by a coordinate transformation
 \beq
  F_{rt}= \frac{\partial x^{\mu^{\rq}}}{\partial r}\frac{\partial x^{\nu^{\rq}}}{\partial t}F_{\mu^{\rq}\nu^{\rq}}= \frac{\partial r^{\rq}}{\partial r}\frac{\partial t^{\rq}}{\partial t}F_{r^{\rq}t^{\rq}}-\frac{\partial t^{\rq}}{\partial r}\frac{\partial r^{\rq}}{\partial t}F_{r^{\rq}t^{\rq}}.
  \eeq
  This gives
 \beq
 F_{rt}= \frac{\gamma Q\Hat{x}}{r^{{\rq}^2}}-\frac{\left(\gamma-1\right)Q}{\beta^2}\frac{\Vec{\beta}\cdot\Hat{x}}{r^{{\rq}^2}}\Vec{\beta},
 \eeq
 or
 \beq
 F_{rt}= \frac{\gamma Q\Vec{x^{\rq}}}{r^{{\rq}^3}}-\frac{\left(\gamma-1\right)Q}{\beta^2}\frac{\Vec{\beta}\cdot\Vec{x^{\rq}}}{r^{{\rq}^3}}\Vec{\beta}.
 \eeq
 Let the velocity $\Vec{\beta}$ be in the $X$ direction. Then
 \beq\label{eq:x}
 x^{\rq}=\gamma\left(x-\beta t\right), \quad y^{\rq}=y, \quad z^{\rq}=z,
 \eeq
 and the radial distance
 \beq\label{eq:r}
r^{\rq}=\left(\gamma^{2}\left(x-\beta t\right)^{2}+y^2+z^2\right)^\frac{1}{2}= \left(\gamma^2\left(t-\beta x\right)^2-t^2+r^2\right)^\frac{1}{2}.
 \eeq
  Generalizing eq. \eqref{eq:x} and eq. \eqref{eq:r} when velocity $\Vec{\beta}$ is in arbitrary direction
   \beq
 \Vec{x^{\rq}}=\Vec{x}+ \frac{\gamma-1}{\beta^2}\left(\Vec{\beta}\cdot\Vec{x}\right)\Vec{\beta}-\gamma\Vec{\beta}t 
 \eeq
 \beq
r^{\rq}=\left(\gamma^2\left(t-r\Hat{x}\cdot\Vec{\beta}\right)^2-t^2+r^2\right)^\frac{1}{2}.
 \eeq
We get the radial component $F_{rt}$ of the field strength  
 \beq
 F_{rt}=\frac{Q\gamma \left(r-t\Hat{x}\cdot \Vec{\beta}\right)}{\left(\gamma^2\left(t-r\Hat{x}\cdot\Vec{\beta}\right)^2-t^2+r^2\right)^\frac{3}{2}}.
\eeq 
Using the retarded null coordinates $u=t-r$ and taking limit $r\rightarrow \infty$, holding $u$ constant to reach $\mathscr{I^{+}}$
\beq
\label{scri+}
F_{ru}|_{\mathscr{I^{+}}}=\frac{Q}{r^2\gamma^2\left(1-\Hat{x} \cdot \Vec{\beta}\right)^2}.
\eeq
To find the field strength at $\mathscr{I^{-}}$, we use advanced coordinates $v=t+r$, holding $v$ constant and taking $r \rightarrow \infty$ limit
\beq
\label{scri-}
F_{rv}|_{\mathscr{I^{-}}}=\frac{Q}{r^2\gamma^2\left(1+\Hat{x} \cdot \Vec{\beta}\right)^2}.
\eeq
\subsection{Antipodal matching condition}

Now, eq. \eqref{scri+} does not depend on $u$, taking $u \to -\infty$ to reach $\mathscr{I}_-^+$ leaves the expression eq.\eqref{scri+} unchanged. 
We get
\beq
\label{scri+-}
F_{ru}|_{\mathscr{I}_-^+}=\frac{Q}{r^2\gamma^2\left(1-\Hat{x} \cdot \Vec{\beta}\right)^2}.
\eeq
Also, eq. \eqref{scri-} does not depend on $v$, taking $v \to \infty$ to reach $\mathscr{I}_-^+$ leaves the expression eq.\eqref{scri-} unchanged. We get
\beq
\label{scri-+}
F_{rv}|_{\mathscr{I}_+^-}=\frac{Q}{r^2\gamma^2\left(1+\Hat{x} \cdot \Vec{\beta}\right)^2}.
\eeq
We note that
\begin{equation}
\begin{split}
&F_{ru}|_{\mathscr{I^{+}}}=F_{ru}|_{\mathscr{I}_-^+}=\frac{Q}{r^2\gamma^2\left(1-\Hat{x} \cdot \Vec{\beta}\right)^2}\\
&F_{rv}|_{\mathscr{I^{-}}}=F_{rv}|_{\mathscr{I}_+^-}=\frac{Q}{r^2\gamma^2\left(1+\Hat{x} \cdot \Vec{\beta}\right)^2}.
\end{split}
\end{equation}

We have the following antipodal matching  condition for the leading part of field strength between in the limit $r\rightarrow \infty$\cite{Strominger:2017zoo}
\beq
 \lim_{r \to \infty}r^2F_{ru}\left(\Hat{x}\right)|_{\mathscr{I^{+}_{-}}}= \lim_{r \to \infty}r^2F_{rv}\left(-\Hat{x}\right)|_{\mathscr{I^{-}_{+}}}.
\eeq
An interesting feature of the Li\'enard-Wiechert field is that the result depends on the order in which we take limits when moving from the bulk to asymptotic regions. Specifically, taking the limit to $\mathscr{I}^+$ (future null infinity) first and then to $i^0$ (spatial infinity) produces a different result than taking the limit to $\mathscr{I}^-$ (past null infinity) first and then to $i^0$.

The solution shows distinct values at fixed angles on $\mathscr{I}_+^+$ and $\mathscr{I}_+^-$, yet maintains an antipodal matching condition. This means that while the values differ at corresponding points on $\mathscr{I}^+$ and $\mathscr{I}^-$ for the same angle $\hat{x}$, they are related through a transformation that maps each point on the celestial sphere to its antipodal point.

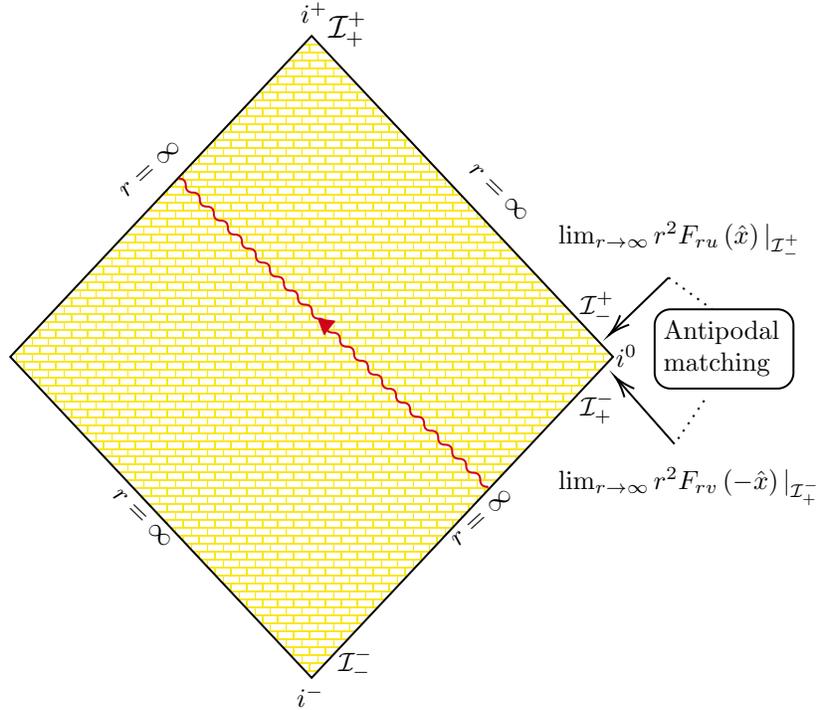
\begin{figure}[H]
	\centering
\tikzset{every picture/.style={line width=0.75pt}} 

\begin{tikzpicture}[x=0.75pt,y=0.75pt,yscale=-1,xscale=1]

\draw  [pattern=_7q7remqtl,pattern size=6pt,pattern thickness=0.75pt,pattern radius=0pt, pattern color={rgb, 255:red, 248; green, 231; blue, 28}] (332,63) -- (484,225) -- (332,387) -- (180,225) -- cycle ;
\draw [color={rgb, 255:red, 208; green, 2; blue, 27 }  ,draw opacity=1 ]   (265,135) .. controls (267.36,135) and (268.54,136.18) .. (268.54,138.54) .. controls (268.54,140.89) and (269.72,142.07) .. (272.07,142.07) .. controls (274.43,142.07) and (275.61,143.25) .. (275.61,145.61) .. controls (275.61,147.96) and (276.79,149.14) .. (279.14,149.14) .. controls (281.5,149.14) and (282.68,150.32) .. (282.68,152.68) .. controls (282.68,155.03) and (283.86,156.21) .. (286.21,156.21) .. controls (288.57,156.21) and (289.75,157.39) .. (289.75,159.75) .. controls (289.75,162.1) and (290.93,163.28) .. (293.28,163.28) .. controls (295.64,163.28) and (296.82,164.46) .. (296.82,166.82) .. controls (296.82,169.18) and (298,170.36) .. (300.36,170.36) .. controls (302.71,170.36) and (303.89,171.54) .. (303.89,173.89) .. controls (303.89,176.25) and (305.07,177.43) .. (307.43,177.43) .. controls (309.78,177.43) and (310.96,178.61) .. (310.96,180.96) .. controls (310.96,183.32) and (312.14,184.5) .. (314.5,184.5) .. controls (316.85,184.5) and (318.03,185.68) .. (318.03,188.03) .. controls (318.03,190.39) and (319.21,191.57) .. (321.57,191.57) .. controls (323.92,191.57) and (325.1,192.75) .. (325.1,195.1) .. controls (325.1,197.46) and (326.28,198.64) .. (328.64,198.64) .. controls (331,198.64) and (332.18,199.82) .. (332.18,202.18) .. controls (332.18,204.53) and (333.36,205.71) .. (335.71,205.71) .. controls (338.07,205.71) and (339.25,206.89) .. (339.25,209.25) .. controls (339.25,211.6) and (340.43,212.78) .. (342.78,212.78) .. controls (345.14,212.78) and (346.32,213.96) .. (346.32,216.32) .. controls (346.32,218.67) and (347.5,219.85) .. (349.85,219.85) .. controls (352.21,219.85) and (353.39,221.03) .. (353.39,223.39) .. controls (353.39,225.74) and (354.57,226.92) .. (356.92,226.92) .. controls (359.28,226.92) and (360.46,228.1) .. (360.46,230.46) .. controls (360.46,232.81) and (361.64,233.99) .. (363.99,233.99) .. controls (366.35,233.99) and (367.53,235.17) .. (367.53,237.53) .. controls (367.53,239.89) and (368.71,241.07) .. (371.07,241.07) .. controls (373.42,241.07) and (374.6,242.25) .. (374.6,244.6) .. controls (374.6,246.96) and (375.78,248.14) .. (378.14,248.14) .. controls (380.49,248.14) and (381.67,249.32) .. (381.67,251.67) .. controls (381.67,254.03) and (382.85,255.21) .. (385.21,255.21) .. controls (387.56,255.21) and (388.74,256.39) .. (388.74,258.74) .. controls (388.74,261.1) and (389.92,262.28) .. (392.28,262.28) .. controls (394.63,262.28) and (395.81,263.46) .. (395.81,265.81) .. controls (395.81,268.17) and (396.99,269.35) .. (399.35,269.35) .. controls (401.71,269.35) and (402.89,270.53) .. (402.89,272.89) .. controls (402.89,275.24) and (404.07,276.42) .. (406.42,276.42) .. controls (408.78,276.42) and (409.96,277.6) .. (409.96,279.96) .. controls (409.96,282.31) and (411.14,283.49) .. (413.49,283.49) .. controls (415.85,283.49) and (417.03,284.67) .. (417.03,287.03) .. controls (417.03,289.38) and (418.21,290.56) .. (420.56,290.56) -- (421,291) -- (421,291) ;
\draw  [color={rgb, 255:red, 208; green, 2; blue, 27 }  ,draw opacity=1 ][fill={rgb, 255:red, 208; green, 2; blue, 27 }  ,fill opacity=1 ] (338.57,213.77) -- (335.71,206.18) -- (343.76,207.19) ;
\draw    (512,185) -- (483.44,212.61) ;
\draw [shift={(482,214)}, rotate = 315.97] [color={rgb, 255:red, 0; green, 0; blue, 0 }  ][line width=0.75]    (10.93,-3.29) .. controls (6.95,-1.4) and (3.31,-0.3) .. (0,0) .. controls (3.31,0.3) and (6.95,1.4) .. (10.93,3.29)   ;
\draw    (515,269) -- (486.32,236.5) ;
\draw [shift={(485,235)}, rotate = 48.58] [color={rgb, 255:red, 0; green, 0; blue, 0 }  ][line width=0.75]    (10.93,-3.29) .. controls (6.95,-1.4) and (3.31,-0.3) .. (0,0) .. controls (3.31,0.3) and (6.95,1.4) .. (10.93,3.29)   ;
\draw  [dash pattern={on 0.84pt off 2.51pt}]  (512,185) -- (534,201) ;
\draw  [dash pattern={on 0.84pt off 2.51pt}]  (530,247) -- (515,269) ;
\draw   (505,209) .. controls (505,204.58) and (508.58,201) .. (513,201) -- (567,201) .. controls (571.42,201) and (575,204.58) .. (575,209) -- (575,233) .. controls (575,237.42) and (571.42,241) .. (567,241) -- (513,241) .. controls (508.58,241) and (505,237.42) .. (505,233) -- cycle ;

\draw (414.79,124.81) node [anchor=north west][inner sep=0.75pt]  [font=\footnotesize,rotate=-44.07]  {$r=\infty $};
\draw (235.13,289.05) node [anchor=north west][inner sep=0.75pt]  [font=\footnotesize,rotate=-44.07]  {$r=\infty $};
\draw (399.89,316.47) node [anchor=north west][inner sep=0.75pt]  [font=\footnotesize,rotate=-317.61]  {$r=\infty $};
\draw (232.66,140.39) node [anchor=north west][inner sep=0.75pt]  [font=\footnotesize,rotate=-317.61]  {$r=\infty $};
\draw (325,44.4) node [anchor=north west][inner sep=0.75pt]  [font=\footnotesize]  {$i^{+}$};
\draw (323,389.4) node [anchor=north west][inner sep=0.75pt]  [font=\footnotesize]  {$i^{-}$};
\draw (466,190.4) node [anchor=north west][inner sep=0.75pt]  [font=\footnotesize]  {$\mathscr{I}_{-}^{+}$};
\draw (466,241.4) node [anchor=north west][inner sep=0.75pt]  [font=\footnotesize]  {$\mathscr{I}_{+}^{-}$};
\draw (344,370.4) node [anchor=north west][inner sep=0.75pt]  [font=\footnotesize]  {$\mathscr{I}_{-}^{-}$};
\draw (339,49.4) node [anchor=north west][inner sep=0.75pt]    {$\mathscr{I}_{+}^{+}$};
\draw (455,155.4) node [anchor=north west][inner sep=0.75pt]  [font=\footnotesize]  {$\lim_{r \to \infty}r^2F_{ru}\left(\Hat{x}\right)|_{\mathscr{I^{+}_{-}}}$};
\draw (455,279.4) node [anchor=north west][inner sep=0.75pt]  [font=\footnotesize]  {$\lim_{r \to \infty}r^2F_{rv}\left(-\Hat{x}\right)|_{\mathscr{I^{-}_{+}}}$};
\draw (508,205) node [anchor=north west][inner sep=0.75pt]  [font=\footnotesize] [align=left] {Antipodal \\matching};
\draw (484,217.4) node [anchor=north west][inner sep=0.75pt]  [font=\footnotesize]  {$i^{0}$};

\end{tikzpicture}

	\caption{Penrose diagram of flat space: past and future null infinities are $\mathscr{I}^{\pm}$. The points $i^{\pm}$ represent past and future timelike infinity. Antipodal matching of flat-space Li\'enard-Wiechert fields $\lim_{r \to \infty}r^2F_{ru}\left(\Hat{x}\right)|_{\mathscr{I^{+}_{-}}}= \lim_{r \to \infty}r^2F_{rv}\left(-\Hat{x}\right)|_{\mathscr{I^{-}_{+}}}.$ }
	\label{fig:antipodalflat}
\end{figure}

The flat space metric in double null coordinates $(u,v,z, \bar{z})$ can be written as
 \beq
 ds^2=-du\,dv+2\left(\frac{v-u}{2}\right)^{2}\gamma_{z\bar{z}}\,dz\, d\bar{z}.
 \eeq
 In the special double null gauge developed in \cite{Krishnan:2021nqo, Krishnan:2022eff,Krishnan:2023zdd}, the antipodal matching becomes \footnote{We should have used a $\tilde{}$ to distinguish between field strengths in two gauges but it will cause unnecessary clutter.}
  \beq
  \lim_{v \to \infty}v^{2}F_{uv}|_{\mathscr{I_{-}^{+}}}\left(\Hat{x}\right)=\lim_{u \to -\infty}u^{2}F_{uv}|_{\mathscr{I_{+}^{-}}}\left(-\Hat{x}\right).
  \eeq
We draw the fig. \ref{fig:antipodalflat} for illustration. In the fig. \ref{fig:antipodalflat}, the past and future null infinities are $\mathscr{I}^{\pm}$. The points $i^{\pm}$ represent past and future timelike infinity. 

\subsection{Li\'enard-Wiechert potentials and fields in AdS}
 The potential for a static charge at origin in AdS space in global coordinates is \cite{Hijano:2020szl}
 \beq
 A_{\mu}=Q \cot \rho\hspace{0.1cm} \delta_{\mu}^{\tau}.
 \eeq
The field strength corresponding to a static charge at origin is
\beq
F_{\rho\tau}=\nabla_{\rho}A_{\tau}-\nabla_{\tau}A_{\rho}=\left(\partial_{\rho}-\tan\rho\right)A_{\tau}=-\frac{Q}{\sin^2\rho}\left(\cos\rho+\sin^2\rho\right).
\eeq
As we did in the previous section, we will use the isometries of AdS to transition from the rest frame of a charged particle to a boosted frame. This process parallels how we apply Lorentz boosts in flat spacetime to shift between frames of reference. The isometry group of the Lorentzian $\mathrm{AdS}_{4}$ is the $SO(3,2)$ group. It consists of
\begin{itemize}
    \item  spatial rotations $SO(3)$
    \item  rotations in the $X^1- X^2$ plane
    \item  boosts with respect to $X^1$ and $X^2$ time directions.
\end{itemize}  
The generators of spatial rotations and boosts with respect to $X^2$ time belong to $so(3,1)$ subalgebra while those of rotations in the $X^1-X^2$ plane and boosts with respect to the $X^1$ time are specific to $so(3,2)$ algebra\cite{Cotaescu:2017ywe}. There are 10 conserved quantities along timelike geodesics associated to the killing vectors of the $SO(3,2)$ group 
\beq
k_{(AB)\mu}=X_A\, \partial_{\mu}X_B-X_B\,\partial_{\mu}X_A, \qquad X_A=\eta_{AB}X^B.
\eeq
We consider a rotation in the $X^1-X^2$ plane
\begin{align}
X^1\rightarrow X^1\cos\alpha-X^2\sin\alpha \\
X^2\rightarrow X^1\sin\alpha+X^2\cos\alpha.
\end{align}
In global coordinates it can be easily seen that this amounts to a translation of the global time, $\tau \rightarrow \tau+\alpha$.
Moreover a  boost in $X^2$ time direction and $X^3$ spatial direction is
\begin{align}
X^1\rightarrow X^1\\ \nonumber
X^2 \rightarrow X^2\cosh\alpha+X^3\sinh\alpha\\ \nonumber
X^3 \rightarrow X^3\cosh\alpha+X^2\sinh\alpha\\ \nonumber
X^4\rightarrow X^4\\ \nonumber
X^5\rightarrow X^5  \nonumber
\end{align}
where $\alpha$ is the boost parameter. In global coordinates, this implies
\begin{align}
\tau \rightarrow \arctan\left(\tan\tau\cosh\alpha+\sin\rho\sec\tau\sin\theta\cos\phi\sinh\alpha\right),\\
\rho \rightarrow \arccos\left(\frac{L}{\sqrt{(X^1)^2+\left(X^2\cosh\alpha+X^3\sinh\alpha\right)^2}}\right).
\end{align}
where $X^i$ are given in eq.\eqref{eq:embeddingglobal}.
Similarly we can check the transformations of global coordinates for boosts with respect to the $X^1$ time direction. Now take the AdS origin as a fixed point, in embedding coordinates it is $P_0=(L,0,0,0,0)$. There exist a non unique subgroup of $SO(3,2)$ which can move this point around over all of AdS i.e., there exist a transformation $\Lambda \in SO(3,2)$ such that
\beq\label{eq:transformatiom}
X(\tau,\rho,\theta,\phi)=\Lambda(\tau,\rho,\theta,\phi)P_0.
\eeq
We consider the ``inertial" frames in AdS space as follows
\begin{itemize}
    \item S: Frame in which particle is at rest at origin.
    \item S': Rest frame of the observer.
\end{itemize}

At a common initial time, let the origins of frames S and S' coincide and their clocks be synchronized. This synchronization is essential for comparing measurements in both frames. An observer at rest in frame S sees an observer at rest in frame S' moving on a timelike geodesic, with velocity in the direction opposite to that of a charged particle. We can choose an arbitrary point in both frames
\beq
X\left(\tau,\rho,\theta,\phi\right)=\left(L\frac{\cos\tau}{\cos\rho},L\frac{\sin\tau}{\cos\rho},L\tan\rho\sin\theta\cos\phi,L\tan\rho\sin\theta\sin\phi,L\tan\rho\cos\theta\right),
\eeq
\beq
X^{\rq}\left(\tau^{\rq},\rho^{\rq},\theta^{\rq},\phi^{\rq}\right)=\left(L\frac{\cos\tau^{\rq}}{\cos\rho^{\rq}},L\frac{\sin\tau^{\rq}}{\cos\rho^{\rq}},L\tan\rho^{\rq}\sin\theta^{\rq}\cos\phi^{\rq},L\tan\rho^{\rq}\sin\theta^{\rq}\sin\phi^{\rq},L\tan\rho^{\rq}\cos\theta^{\rq}\right).
\eeq
The above transformation eq.\eqref{eq:transformatiom} allows us to reach both these points from $P_0$ as $X=\Lambda P_0$ and $X^{\rq}=\Lambda^{\rq} P_0$. We now can define $\Lambda^{\ast}$ as $\Lambda^{\ast}=\Lambda^{\rq}\Lambda^{-1}$ such that 
\beq
X^{\rq}=\Lambda^{\ast}X, \quad \Lambda^{\ast} \in SO(3,2).
\eeq
 The condition of the two frames coinciding at a common initial time implies that $\Lambda^{\ast}$ turns out to be usual ``Lorentz'' transformations, i.e., boosts for $X^2$ time direction. The transformation matrix in terms of energy and momenta of the particle is given by 
\begin{equation}
\Lambda^{\ast}=
\begin{bmatrix}
1 & 0 & 0 & 0 & 0 \\
0 & \frac{\omega_{\Vec{p}}}{m} & \frac{p^{1}}{m} & \frac{p^{2}}{m} & \frac{p^{3}}{m}\\
0 & \frac{p^{1}}{m} & 1+ \left(\frac{\omega_{\Vec{p}}}{m}-1\right)\frac{(p^{{1}})^{2}}{\Vec{{p}^2}} & \left(\frac{\omega_{\Vec{p}}}{m}-1\right)\frac{p^{1}p^{2}}{\Vec{{p}^2}} & \left(\frac{\omega_{\Vec{p}}}{m}-1\right)\frac{p^{1}p^{3}}{\Vec{{p}^2}}\\
0 & \frac{p^{2}}{m} & \left(\frac{\omega_{\Vec{p}}}{m}-1\right)\frac{p^{1}p^{2}}{\Vec{{p}^2}} & 1+\left(\frac{\omega_{\Vec{p}}}{m}-1\right)\frac{(p^{{2}})^2}{\Vec{{p}^2}} & \left(\frac{\omega_{\Vec{p}}}{m}-1\right)\frac{p^{2}p^{3}}{\Vec{{p}^2}}\\
0 & \frac{p^{3}}{m} & \left(\frac{\omega_{\Vec{p}}}{m}-1\right)\frac{p^{1}p^{3}}{{\Vec{p}^2}}  & \left(\frac{\omega_{\Vec{p}}}{m}-1\right)\frac{p^{2}p^{3}}{\Vec{{p}^2}} & 1+\left(\frac{\omega_{\Vec{p}}}{m}-1\right)\frac{(p^{{3}})^2}{\Vec{{p}^2}} 
\end{bmatrix}.
\end{equation}
The new $(\tau,\rho)$ coordinates are related to old ones as follows
\beq
\tan\tau=\frac{\omega_{\Vec{p}}}{m}\tan\tau^{\rq}-\frac{|\Vec{p}|\hspace{0.05cm}\sin\rho^{\rq}\Omega^{\rq}.\Hat{p}}{m\cos\tau^{\rq}}, \label{eq:tau}
\eeq
\beq\label{eq:rho}
\cos\rho=\frac{\cos\rho^{\rq}}{\cos\tau^{\rq}}\frac{1}{\sqrt{1+\tan^2\tau}}. 
\eeq
where $\hat{p}$ is the direction of the charged particle. The electromagnetic field strength in the boosted frame can be obtained by a coordinate transformation
\beq
F_{\rho^{\rq}\tau^{\rq}}= \left(\frac{\partial\rho}{\partial\rho^{\rq}}\frac{\partial\tau}{\partial\tau^{\rq}}-\frac{\partial\rho}{\partial\tau^{\rq}}\frac{\partial\tau}{\partial\rho^{\rq}}\right)F_{\rho\tau}.
\eeq
The explicit expression for the field strength of a boosted charge in AdS in global coordinates is
\begin{equation}\label{eq:bulkstrength}
\small 
\begin{split}
F_{\rho^{\rq}\tau^{\rq}}&=\frac{-1}{\sqrt{1- \cos^2\rho}}\frac{1}{1+\tan^2\tau} \\
&\left[\left(-\sec\tau^{\rq}\left\lbrace\frac{-\sin\rho^{\rq}}{\sqrt{1+\tan^2\tau}}+\frac{\tan\tau}{\left(1+\tan^2\tau\right)^{\frac{3}{2}}}\left(\frac{-|\Vec{p}|\cos^2\rho^{\rq}\Omega^{\rq}\cdot\Hat{p}}{m\cos\tau^{\rq}}\right)\right\rbrace\right)\right.\\
&\left(\frac{\omega_{\Vec{p}}}{m}\sec^2\tau^{\rq}-\frac{|\Vec{p}|\sin\rho^{\rq}\Omega^{\rq}\cdot\Hat{p}\sec\tau^{\rq}\tan\tau^{\rq}}{m}\right) \\ 
&-\left.\left(\frac{|\Vec{p}|\cos^2\rho^{\rq}\Omega^{\rq}\cdot\Hat{p}}{m\cos\tau^{\rq}}\right)\left\lbrace\frac{\sec\tau^{\rq}\tan\tau^{\rq}}{\sqrt{1+\tan^2\tau}}+\tan\tau\sec\tau^{\rq}\left(\frac{\omega_{\Vec{p}}}{m}\sec^2\tau^{\rq}-\frac{|\Vec{p}|\sin\rho^{\rq}\Omega^{\rq}\cdot\Hat{p}\sec\tau^{\rq}\tan\tau^{\rq}}{m}\right)\right\rbrace\right],
\end{split}
\end{equation}
where $\tan\tau$ and $\cos\rho$ are given by eq. \eqref{eq:tau} and \eqref{eq:rho}. 
\subsection{Antipodal matching condition in the flat limit of AdS}

The expression for the field strength simplifies greatly when we go to the boundary of AdS along null geodesics. For example, we have the following limits
\beq
\lim_{\rho^{\rq} \to \frac{\pi}{2}, \tau^{\rq} \to \frac{\pi}{2}}F_{\rho^{\rq}\tau^{\rq}}=\frac{Q}{\gamma^2\left(1-\Vec{\beta} \cdot \Hat{x^{\rq}}\right)^2}, \hspace{0.5cm}  \lim_{\rho^{\rq} \to \frac{\pi}{2}, \tau^{\rq} \to -\frac{\pi}{2}}F_{\rho^{\rq}\tau^{\rq}}=\frac{Q}{\gamma^2\left(1+\Vec{\beta} \cdot \Hat{x^{\rq}}\right)^2},
\eeq
where we have defined
\beq
\gamma=\frac{\omega_{\Vec{p}}}{m}, \hspace{1.0cm} \gamma \Vec{\beta}=\frac{\Vec{p}}{m}, \qquad \Vec{p}=|\Vec{p}|\hat{p}, \qquad \Omega' \equiv \hat{x^{\rq}}
\eeq
to emphasize the similarities with the flat space case. Moreover we have
\beq
\lim_{\rho^{\rq} \to \frac{\pi}{2}, \tau^{\rq} \to \frac{\pi}{2}}\frac{1}{\cos\rho^{\rq}}A_{\tau^{\rq}}=\frac{Q}{\gamma^2\left(1-\Vec{\beta} \cdot \Hat{x^{\rq}}\right)^2}, \hspace{0.5cm}  \lim_{\rho^{\rq} \to \frac{\pi}{2}, \tau^{\rq} \to -\frac{\pi}{2}}\frac{1}{\cos \rho^{\rq}}A_{\tau^{\rq}}=\frac{Q}{\gamma^2\left(1+\Vec{\beta} \cdot \Hat{x^{\rq}}\right)^2}.
\eeq 
The specific limits $\tau \rightarrow \frac{\pi}{2}$ and $\tau \rightarrow -\frac{\pi}{2}$ are taken because small fringes about these global times play the role of ${\mathscr{I^{+}}}$ and ${\mathscr{I^{-}}}$ in the boundary CFT \cite{Hijano:2020szl}. Further de Gioia and Raclariu have shown that in the strips about these global times, the conformal group of a 3-dimensional $\mathrm{CFT}_3$ on the Lorentzian cylinder (holographic or not) gets enhanced to an infinitesimal symmetry in large radius limit, and the vector fields generating this symmetry obey the extended $\mathrm{BMS}_{4}$ algebra after an \text{In\"{o}n\"{u}--Wigner} like contraction, see e.g., \cite{deGioia:2022fcn,deGioia:2023cbd}.

\begin{figure}[H]
	\centering
\tikzset{every picture/.style={line width=0.75pt}} 

\begin{tikzpicture}[x=0.75pt,y=0.75pt,yscale=-1,xscale=1]

\draw  [color={rgb, 255:red, 74; green, 144; blue, 226 }  ,draw opacity=1 ] (425,146.15) -- (425,324.85) .. controls (425,339.84) and (384.48,352) .. (334.5,352) .. controls (284.52,352) and (244,339.84) .. (244,324.85) -- (244,146.15) .. controls (244,131.16) and (284.52,119) .. (334.5,119) .. controls (384.48,119) and (425,131.16) .. (425,146.15) .. controls (425,161.14) and (384.48,173.3) .. (334.5,173.3) .. controls (284.52,173.3) and (244,161.14) .. (244,146.15) ;
\draw  [color={rgb, 255:red, 74; green, 144; blue, 226 }  ,draw opacity=1 ][line width=2.25]  (244,322.46) .. controls (244,306.15) and (284.52,292.93) .. (334.5,292.93) .. controls (384.48,292.93) and (425,306.15) .. (425,322.46) .. controls (425,338.78) and (384.48,352) .. (334.5,352) .. controls (284.52,352) and (244,338.78) .. (244,322.46) -- cycle ;
\draw  [color={rgb, 255:red, 208; green, 2; blue, 27 }  ,draw opacity=1 ][line width=2.25]  (244,146.15) .. controls (244,131.16) and (284.52,119) .. (334.5,119) .. controls (384.48,119) and (425,131.16) .. (425,146.15) .. controls (425,161.14) and (384.48,173.3) .. (334.5,173.3) .. controls (284.52,173.3) and (244,161.14) .. (244,146.15) -- cycle ;

\draw (198,132.4) node [anchor=north west][inner sep=0.75pt]  [font=\footnotesize]  {$\tau =\cfrac{\pi }{2}$};
\draw (192,312.4) node [anchor=north west][inner sep=0.75pt]  [font=\footnotesize]  {$\tau =-\cfrac{\pi }{2}$};

\end{tikzpicture}

	\caption{The red fringe about global time $\tau=\frac{\pi}{2}$ is antipodally identified with the blue fringe at $\tau=-\frac{\pi}{2}$. In the flat limit, these fringes resemble future and past null infinity respectively. The boundary region between $\tau=-\frac{\pi}{2}$ and $\tau=\frac{\pi}{2}$ can be thought of as spatial infinity in large $L$ limit.}
	\label{fig:antipodalflatlimit}
\end{figure}
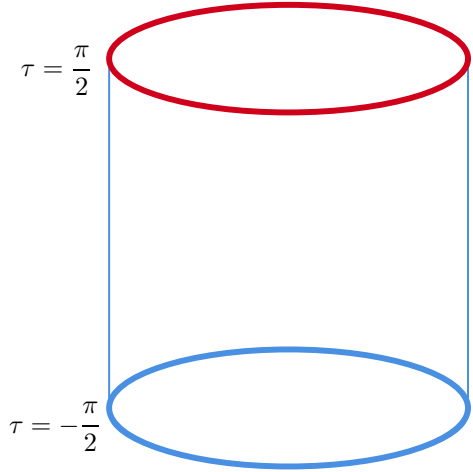

The following coordinate change in the limit $L\rightarrow \infty$ gives the Lorentz boost in flat space in spherical coordinates\\
$$\tau=\frac{t}{L}, \qquad \tan\rho=\frac{r}{L}$$
and the transformation matrix $\Lambda^{\ast}$ reduces to the usual flat space Lorentz transformations and we retrieve the flat space antipodal matching of Li\'enard-Wiechert solution, the antipodal matching was previously observed in flat space in \cite{Strominger:2017zoo}.
We illustrate the antipodal matching of fields in the flat limit of AdS. See fig. \ref{fig:antipodalflatlimit}.

It was previously noted in \cite{Hamilton:2006az} that the antipodal matching condition in global coordinates is  
$$\rho~~\text{invariant},~~\tau \to \tau+\pi,~~\hat{x} \to -\hat{x}.$$
We have explicitly checked the antipodal matching condition holds for our bulk expression for field strength eq.\eqref{eq:bulkstrength}.  

In the flat limit of AdS we get
\beq
\lim_{\rho^{\rq} \to \frac{\pi}{2}, \tau^{\rq} \to \frac{\pi}{2}}F_{\rho^{\rq}\tau^{\rq}}=\frac{Q}{\gamma^2\left(1-\Vec{\beta} \cdot \Hat{x^{\rq}}\right)^2}, \hspace{0.5cm}  \lim_{\rho^{\rq} \to \frac{\pi}{2}, \tau^{\rq} \to -\frac{\pi}{2}}F_{\rho^{\rq}\tau^{\rq}}=\frac{Q}{\gamma^2\left(1+\Vec{\beta} \cdot \Hat{x^{\rq}}\right)^2}.
\eeq
We note that
\begin{equation}
\begin{split}
&\lim_{\rho^{\rq} \to \frac{\pi}{2}, \tau^{\rq} \to \frac{\pi}{2}}F_{\rho^{\rq}\tau^{\rq}}=\lim_{r \to \infty}r^2F_{ru}\left(\Hat{x}\right)|_{\mathscr{I^{+}_{-}}}\\
&\lim_{\rho^{\rq} \to \frac{\pi}{2}, \tau^{\rq} \to -\frac{\pi}{2}}F_{\rho^{\rq}\tau^{\rq}}=\lim_{r \to \infty}r^2F_{rv}\left(\Hat{x}\right)|_{\mathscr{I^{-}_{+}}}.
\end{split}
\end{equation}
Now, in the flat limit of AdS the antipodal matching condition becomes 
\beq
 \lim_{\rho^{\rq} \to \frac{\pi}{2}, \tau^{\rq} \to \frac{\pi}{2}}F_{\rho^{\rq}\tau^{\rq}}\left(\Hat{x}\right)=\lim_{\rho^{\rq} \to \frac{\pi}{2}, \tau^{\rq} \to -\frac{\pi}{2}}F_{\rho^{\rq}\tau^{\rq}}\left(-\Hat{x}\right).
\eeq
To summarize, the component \(F_{ru}\) evaluated on \(\mathscr{I}^+\) (denoted \(F_{ru}|_{\mathscr{I}^+}\)) remains unchanged when further taking the limit to \(\mathscr{I}_-^+\). This is because the expression for \(F_{ru}\) on \(\mathscr{I}^+\) is independent of the retarded time \(u\), allowing us to directly extrapolate
     \[
     F_{ru}\big|_{\mathscr{I}^+} = F_{ru}\big|_{\mathscr{I}_-^+}.
     \]
Similarly, the component \(F_{rv}\) evaluated on \(\mathscr{I}^-\) (denoted \(F_{rv}|_{\mathscr{I}^-}\)) remains unchanged when taking the limit to \(\mathscr{I}_+^-\). The expression for \(F_{rv}\) on \(\mathscr{I}^-\) is independent of the advanced time \(v\), enabling the extrapolation
     \[
     F_{rv}\big|_{\mathscr{I}^-} = F_{rv}\big|_{\mathscr{I}_+^-}.
     \]

Despite these extrapolations, the solution is different at fixed angles on \(\mathscr{I}_+^+\) and \(\mathscr{I}_+^-\). However, these are related through an antipodal matching condition. This means that the field at a point on \(\mathscr{I}_+^+\) with direction \(\hat{x}\) corresponds to the field at the antipodal point on \(\mathscr{I}_+^-\) with direction \(-\hat{x}\).

When considering the flat limit of AdS space, this extrapolation property becomes particularly useful. By evaluating the field at a point on \(\mathscr{I}^+\), we can extrapolate to obtain the field at the corresponding point on \(\mathscr{I}_-^+\). Similarly, evaluating the field on \(\mathscr{I}^-\) allows extrapolation to \(\mathscr{I}_+^-\). 

The region between $\tau=\frac{\pi}{2}$ and $\tau=-\frac{\pi}{2}$ is the spatial infinity in large $L$ limit, and we show that the fields are antipodally matched between  \( \mathscr{I}^{-}_{+} \) $(u=-\infty, \hat{x})$ and  \( \mathscr{I}^{+}_{-} \) $(v=+\infty, -\hat{x})$.

\section{Conclusions and future directions}
\label{concl}
The flat limit of AdS/CFT is concerned with computing flat space observables from boundary correlators in AdS/CFT in large radius limit. The usual method involves bulk reconstruction where a local operator in the bulk has a CFT representation. In this paper, we took an alternative approach and studied the flat limit of AdS/CFT by looking at geodesics in AdS. We computed geodesics in global AdS and determined their time of travel from the origin to the AdS boundary for both massive and massless particles. For a massless particle (null geodesic), the travel time from the origin to the AdS boundary is
\[
\Delta \tau = \frac{\pi}{2}.
\]
For a massive particle (timelike geodesic), the travel time is
\[
\Delta \tau = \frac{\pi}{2} + \frac{i}{2} \log \left( \frac{\omega_{\vec{p}} + m}{\omega_{\vec{p}} - m} \right),
\]
where \(\omega_{\vec{p}}\) is the energy of the particle and \(m\) is its mass. The time of travel for a massive particle is imaginary meaning that they never really reach the boundary but oscillate indefinitely around the center of AdS. By looking at their trajectory in embedding space, it is clear that timelike geodesics are bounded in a diamond region with the boundary of the region formed by null geodesics. The flat space creation and annihilation operators for a particle propagating in the scattering region at the center of AdS can be constructed directly from the boundary operators as
\begin{equation}
\label{geotratime1}
 \sqrt{2 \omega_{\vec{p}}} \,a_{\text{out}/\text{in}, \vec{p}} \sim \int d\tau \, e^{i \omega_{\vec{p}} L(\tau \mp \text{geodesic travel time})}  \mathcal{O}(\tau, \pm \hat{p}).
\end{equation}
As a simple check, we were successfully able to reproduce $2$-point function using eq.\eqref{geotratime1} for both massless and massive particles.

Next, we used this travel time to take the flat limit of Li\'enard-Wiechert fields in AdS. We considered a static charge in AdS, computed its Li\'enard-Wiechert field, and then applied an AdS isometry so that the geodesic of the particle hits the AdS boundary at the point where an operator must be inserted to get a scattering state in the flat limit of AdS. In the flat limit of AdS, we traced the null geodesic, which hits the AdS boundary at the proper time \(\tau = \frac{\pi}{2}\). We showed that the field at the antipodal point
will be 
at \(\tau = -\frac{\pi}{2}\). In fact, the time slices separated by a global time difference $\Delta \tau=\pi$ are antipodally identified, this was first noted by HKLL \cite{Hamilton:2006az} and argued recently by de Gioia and Raclariu \cite{deGioia:2024yne, deGioia:2022fcn}. 

Now, let us mention an elucidatory remark. We then close with some future directions.
\subsection*{An elucidatory remark}

\textbf{Why boosting in $X^1$ direction does not work.}

Let us consider the Lorentz boosts with respect to the $X^1$ time. It means the global time $\tau$ transforms as
\beq
\cot\tau= \frac{\omega}{m}\cot\tau^{\rq}-\frac{| \Vec{p}|\hspace{0.1cm}\sin\rho^{\rq}\Omega^{\rq}.\Hat{p}}{m\sin\tau^{\rq}}.
\eeq
Thus the boosts with respect to $X^1$ are undefined if the initial point is the AdS origin. If we wanted to start with initial point $(\frac{\pi}{2},0,0,0)$, then the condition of coincident frame at the initial time demands that we translate $\tau$ by $-\frac{\pi}{2}$. This means a rotation by an angle $-\frac{\pi}{2}$ in the $X^1-X^2$ plane but this would take $X^1$ into $X^2$ axis which is nothing new.

In other words, if one has a boosted frame for any ``time" vector lying in the $X^1-X^2$ plane then the condition that two origins coincide initially demands that the boost be accompanied by a rotation in $X^1-X^2$ plane by an angle $\theta$. But
\beq
exp(-i\theta\mathcal{R})\,\mathcal{B}_i\,exp(i\theta\mathcal{R})=\mathcal{B}_2,
\eeq
where $\mathcal{R}$ is the generator of rotations in $X^1-X^2$ plane, $\theta$ the angle between the time vector and $X^2$ axis, and $\mathcal{B}_i$ is a boost for a time vector lying in the Euclidean $X^1-X^2$ plane and $\mathcal{B}_2$ is the usual {\footnote{Has a well defined flat limit.}} ``Lorentzian'' boost in $X^2$.


\noindent\makebox[\linewidth]{\rule{0.9\textwidth}{1pt}} 

\subsection*{Future directions.}
\paragraph{Relating the flat limit of AdS amplitudes to celestial amplitudes.}
Using the mapping between creation/annihilation operators and CFT operators at the AdS boundary from AdS geodesics, we can construct the $\mathcal{S}$-matrix using AdS amplitudes. Celestial amplitudes are flat space scattering amplitudes written in a boost eigenbasis by applying Mellin transforms. These transform covariantly under an $SL(2,C)$ transformation. In \cite{deGioia:2024yne, deGioia:2022fcn}, the celestial amplitudes for massless particles are computed from the AdS amplitudes in the flat limit. It would be interesting to compute the celestial amplitudes for the massive case and connect our approach with the flat limit of AdS amplitudes, see recent works \cite{Iacobacci:2022yjo, Sleight:2023ojm, Iacobacci:2024nhw}. Explicit computations for massive celestial amplitudes have been carried out in \cite{Chang:2023ttm, Liu:2024vmx, Liu:2024lbs, Duary:2024cqb} using the spectral representation technique, which is useful for deriving the conformal block expansion of celestial amplitudes. It would be nice to establish a connection with the conformal block expansion of AdS amplitudes in the flat limit. It would also be nice to establish a connection between celestial \(\mathrm{CFT}_0\) amplitudes developed in \cite{Duary:2022onm, Kapec:2022xjw, Stolbova:2023smk, Stolbova:2023cof, Wang:2025quh} and the flat limit of \(\mathrm{CFT}_1\) correlators \cite{Duary:2023gqg}.

\paragraph{Relating the flat limit of AdS amplitudes to carrollian amplitudes.}
Carrollian CFTs have been proposed as holographic duals with co-dimension one to asymptotically flat spacetimes, in contrast to celestial CFTs, which have co-dimension two. In the flat limit of AdS amplitudes, the crucial input is that the smearing regions of the CFT operators are localized around specific time slices of the boundary when the AdS radius becomes large. Depending on the choice of the smearing regions, the scattering can be either massless or massive. In global AdS, the operators are smeared around particular future and past time slices, given by 
\[
\tau = \pm \frac{\pi}{2} + \mathcal{O}(L)^{-1},
\]
to get massless scattering states, and
\[
\tau = \pm \frac{\pi}{2} \pm \frac{i}{2} \log \left( \frac{\omega_{\vec{p}} + m}{\omega_{\vec{p}} - m} \right)+ \mathcal{O}(L)^{-1},
\]
to get massive scattering states. This suggests that the CFT data required to compute $\mathcal{S}$-matrices for massless, and massive particles are encoded in the region localized around \(\tau = \pm \frac{\pi}{2}\), and $\tau = \pm \frac{\pi}{2} \pm \frac{i}{2} \log \left( \frac{\omega_{\vec{p}} + m}{\omega_{\vec{p}} - m} \right)$. Carrollian amplitudes are the natural objects obtained in the flat limit of AdS amplitudes. The relation between carrollian and AdS amplitudes for massless particles was explored in \cite{Bagchi:2023fbj, Bagchi:2023cen, Alday:2024yyj}.\footnote{Recently, Kim et al.\cite{Kim:2023qbl} derived leading soft photon theorem from a generating functional approach to the $\mathcal{S}$-matrix (also known as the AFS $\mathcal{S}$-matrix after pioneering work of Arefeva, Faddeev and Slavnov\cite{Arefeva:1974jv} in 70s) which is tied to the boundary data and asymptotic symmetries. Further, Kraus and Myers\cite{Kraus:2024gso} showed that in the flat limit that the AdS and AFS generating functionals are equivalent.} Recently \cite{Have:2024dff}, massive carrollian fields were constructed from the blowup of timelike infinity \(i^{\pm}\). Carrollian and celestial amplitudes, and their interconnection, has been explored in \cite{Donnay:2022aba, Bagchi:2022emh, Donnay:2022wvx, Mason:2023mti}. It would be interesting to relate the AdS amplitudes in the flat limit to the carrollian amplitudes for massive particles. 

\paragraph{Scattering in asymptotically flat space having black
holes.}
Asymptotically flat black holes can only arise as the flat limit of AdS geometries having small black holes, where the radius of the black holes does not scale with the AdS length scale. It would be great to calculate CFT correlators and then study AdS geodesics to determine the scattering amplitudes in such a background.





\section*{Acknowledgements}
We are immensely grateful to Chethan Krishnan for numerous discussions. The work of SD is supported by the Shuimu Scholar Program of Tsinghua University. Part of this work was done while SU was a visiting student at Raman Research Institute (RRI). SU thanks RRI for their hospitality and support.

\end{document}